\documentclass[preprint]{aastex631}
\usepackage{mathtools}
\usepackage{soul}
\usepackage{glossaries}

\usepackage{multirow}
\begin{document}

\title{Simulation of high-contrast polarimetric observations of debris disks with the Roman Coronagraph Instrument}

\correspondingauthor{Ramya M Anche}
\email{ramyaanche@arizona.edu}

\author[0000-0002-4989-6253]{Ramya M Anche}
\affiliation{Steward Observatory, University of Arizona, 933N Cherry Avenue, Tucson, Arizona, 85721, USA}

\author[0000-0002-0813-4308]{Ewan Douglas}
\affiliation{Steward Observatory, University of Arizona, 933N Cherry Avenue, Tucson, Arizona, 85721, USA}

\author[0000-0001-9204-2299]{Kian Milani}
\affiliation{Steward Observatory, University of Arizona, 933N Cherry Avenue, Tucson, Arizona, 85721, USA}
\affiliation{James C. Wyant College of Optical Sciences, University of Arizona, Tucson, Arizona, 85721, USA}

\author[0000-0001-5082-7442]{Jaren Ashcraft}
\affiliation{Steward Observatory, University of Arizona, 933N Cherry Avenue, Tucson, Arizona, 85721, USA}
\affiliation{James C. Wyant College of Optical Sciences, University of Arizona, Tucson, Arizona, 85721, USA}

\author[0000-0001-6205-9233]{Maxwell A. Millar-Blanchaer}
\affiliation{Department of Physics, University of California, Santa Barbara, CA, 93106, USA}

\author[0000-0002-1783-8817]{John H Debes}
\affiliation{AURA for ESA, Space Telescope Science Institute, 3700 San Martin Drive, Baltimore, MD 21218, USA}

\author[0000-0001-9325-2511]{Julien Milli}
\affiliation{Universit\'{e} Grenoble Alpes, CNRS, IPAG, 38000 Grenoble, France}

\author[0000-0002-0813-4308]{Justin Hom}
\affiliation{Steward Observatory, University of Arizona, 933N Cherry Avenue, Tucson, Arizona, 85721, USA}

\begin{abstract}
The Nancy Grace Roman Space Telescope Coronagraph Instrument will enable the polarimetric imaging of debris disks and inner dust belts in the optical and near-infrared wavelengths, in addition to the high-contrast polarimetric imaging and spectroscopy of exoplanets. The Coronagraph uses two Wollaston prisms to produce four orthogonally polarized images and is expected to measure the polarization fraction with measurement errors $<$ 3\% per spatial resolution element. To simulate the polarization observations through the Hybrid Lyot Coronagraph (HLC) and Shaped Pupil Coronagraph (SPC), we model disk scattering, the coronagraphic point-response function, detector noise, speckles, jitter, and instrumental polarization and calculate the Stokes parameters. To illustrate the potential for discovery and a better understanding of known systems with both the HLC and SPC modes, we model the debris {disks} around Epsilon Eridani and HR 4796A, respectively. For Epsilon Eridani, using astrosilicates with 0.37$\pm$0.01 as the peak input polarization fraction in one resolution element, we recover the peak disk polarization fraction of 0.33$\pm$0.01. Similarly, for HR 4796A, for a peak input polarization fraction of 0.92$\pm$0.01, we obtain the peak output polarization fraction as 0.80$\pm$0.03. The Coronagraph design meets the required precision, and forward modeling is needed to accurately estimate the polarization fraction.
\end{abstract}


\newacronym{AU}{AU}{astronomical Unit [1.5e11 m]}  
\newacronym{pc}{pc}{parsec}
\newacronym{mas}{mas}{milliarcsecond}
\newacronym{nm}{nm}{nanometer}
\newacronym{CTE}{CTE}{coefficient of thermal expansion}
\newacronym{sqarc}{$as^2$}{square arcsecond}

\newacronym{smc}{SMC}{Small Magellanic Cloud}
\newacronym{lmc}{LMC}{Large Magellanic Cloud}
\newacronym{ism}{ISM}{interstellar medium}
\newacronym{mw}{MW}{Milky Way}
\newacronym{epseri}{$\epsilon$ Eri}{Epsilon Eridani}
\newacronym{EKB}{EKB}{Edgeworth-Kuiper Belt}

\newacronym{CFR}{CFR}{Complete Frequency Redistribution}

\newacronym{nasa}{NASA}{National Aeronautics and Space Agency}
\newacronym{esa}{ESA}{European Space Agency}
\newacronym{omi}{OMI}{\textit{Optical Mechanics Inc.}}
\newacronym{gsfc}{GSFC}{\gls{nasa} Goddard Space Flight Center}
\newacronym{stsci}{STScI}{Space Telescope Science Institute}
\newacronym{nsroc}{NSROC}{\gls{nasa} Sounding Rocket Operations Contract}
\newacronym{wff}{WFF}{\gls{nasa} Wallops Flight Facility}
\newacronym{wsmr}{WSMR}{White Sands Missile Range}

\newacronym{irac}{IRAC}{Infrared Array Camera}
\newacronym[plural=CCDs, firstplural=charge-coupled devices (CCDs)]{ccd}{CCD}{charge-coupled device}
\newacronym[plural=EMCCDs, firstplural=electron multiplying charge-coupled devices (EMCCDs)]{EMCCD}{EMCCD}{electron multiplying charge-coupled device}

\newacronym{DM}{DM}{Deformable Mirror}
\newacronym{MCP}{MCP}{ Microchannel Plate }
\newacronym{ipc}{IPC}{Image Proportional Counter}
\newacronym{cots}{COTS}{Commercial Off-The-Shelf}
\newacronym{ISR}{ISR}{incoherent scatter radar}
\newacronym{atcamera}{AT}{angle tracker}
\newacronym{MEMS}{MEMS}{microelectromechanical systems}
\newacronym{QE}{QE}{quantum efficiency}
\newacronym{RTD}{RTD}{Resistance Temperature Detector}
\newacronym{PID}{PID}{Proportional-Integral-Derivative}
\newacronym{PRNU}{PRNU}{photo response non-uniformity}
\newacronym{DSNU}{PRNU}{dark signal non-uniformity}
\newacronym{CMOS}{CMOS}{complementary metal–oxide–semiconductor}
\newacronym{TRL}{TRL}{technology readiness level}
\newacronym{swap}{SWaP}{Size, Weight, and Power}
\newacronym{ConOps}{ConOps}{concept of operations}
\newacronym{NRE}{NRE}{non-recurring engineering}
\newacronym{CBE}{CBE}{current best estimate}

\newacronym{FOV}{FOV}{field-of-view}
\newacronym{NIR}{NIR}{near-infrared}
\newacronym{PV}{PV}{Peak-to-Valley}
\newacronym{MRF}{MRF}{Magnetorheological finishing}
\newacronym{AO}{AO}{Adaptive Optics}
\newacronym{TTP}{TTP}{tip, tilt, and piston}
\newacronym{FPS}{FPS}{fine pointing system}
\newacronym{SHWFS}{SHWFS}{Shack-Hartmann Wavefront Sensor}
\newacronym{OAP}{OAP}{off-axis parabola}
\newacronym{LGS}{LGS}{laser guide star}
\newacronym{WFCS}{WFCS}{wavefront control system}
\newacronym{OPD}{OPD}{optical path difference}
\newacronym{UA}{UA}{University of Arizona}
\newacronym{MEL}{MEL}{Master Equipment List}
\newacronym{LEO}{LEO}{low-earth orbit}
\newacronym{GEO}{GEO}{geosynchronous orbit}
\newacronym{EFC}{EFC}{electric-field conjugation}
\newacronym{LDFC}{LDFC}{linear dark field control}
\newacronym{DAC}{DAC}{digital-to-analog converter}
\newacronym{FEA}{FEA}{finite element analysis}

\newacronym{SiC}{SiC}{Silicon Carbide}
\newacronym{ESPA}{ESPA}{EELV Secondary Payload Adapter}
\newacronym{EEID}{EEID}{Earth-equivalent Insolation Distance, the distance from the star where the incident energy density is that of the Earth received from the Sun}
\newacronym{LLOWFS}{LLOWFS}{Lyot low-order wavefront sensor}
\newacronym{STOP}{STOP}{Structural-Thermal-Optical-Performance}

\newacronym{resel}{resel}{resolution element}

\newacronym{acs}{ACS}{Attitude Control System}
\newacronym{orsa}{ORSA}{Ogive Recovery System Assembly}
\newacronym{gse}{GSE}{Ground Station Equipment}
\newacronym{FSM}{FSM}{Fast Steering Mirror}

\newacronym{WFS}{WFS}{wavefront sensor}
\newacronym{LSI}{LSI}{Lateral Shearing Interferometer}
\newacronym{VVC}{VVC}{Vector Vortex Coronagraph}
\newacronym{VNC}{VNC}{Visible Nulling Coronagraph}
\newacronym{CGI}{CGI}{Coronagraph Instrument}
\newacronym{IWA}{IWA}{Inner Working Angle}
\newacronym{OWA}{OWA}{Outer Working Angle}
\newacronym{NPZT}{N-PZT}{Nuller piezoelectric transducer}
\newacronym{ZWFS}{ZWFS}{Zernike wavefront sensor}
\newacronym{SPC}{SPC}{Shaped Pupil Coronagraph}
\newacronym{HLC}{HLC}{Hybrid-Lyot Coronagraph}
\newacronym{ADI}{ADI}{angular differential imaging}
\newacronym{RDI}{RDI}{reference differential imaging}
\newacronym{LOWFSC}{LOWFS/C}{low-order wavefront sensing and control}
\newacronym{HOWFSC}{HOWFS/C}{high-order wavefront sensing and control}
\newacronym{WFSC}{WFSC}{wavefront sensing and control}

\newacronym{HST}{HST}{Hubble Space Telescope}
\newacronym{GPS}{GPS}{Global Positioning System}
\newacronym{ISS}{ISS}{International Space Station}
\newacronym[description=Advanced CCD Imaging Spectrometer]{acis}{ACIS}{Advanced \gls{ccd} Imaging Spectrometer}
\newacronym{stis}{STIS}{\textit{Space Telescope Imaging Spectrograph}}
\newacronym{mcp}{MCP}{Microchannel Plate}
\newacronym{jwst}{JWST}{$\textit{James Webb Space Telescope}$}
\newacronym{fuse}{FUSE}{$\textit{FUSE}$}
\newacronym{galex}{GALEX}{$\textit{Galaxy Evolution Explorer}$}
\newacronym{spitzer}{Spitzer}{$\textit{Spitzer Space Telescope}$}
\newacronym{mips}{MIPS}{Multiband Imaging Photometer for \gls{spitzer}}
\newacronym{gissmo}{GISSMO}{Gas Ionization Solar Spectral Monitor}
\newacronym{iue}{IUE}{International Ultraviolet Explorer}
\newacronym{spinr}{SPINR}{$\textit{Spectrograph for Photometric Imaging with Numeric Reconstruction}$}
\newacronym{imager}{IMAGER}{$\textit{Interstellar Medium Absorption Gradient Experiment Rocket}$}
\newacronym{TPF-C}{TPF-C}{Terrestrial Planet Finder Coronagraph}
\newacronym{RAIDS}{RAIDS}{Atmospheric and Ionospheric Detection System }
\newacronym{mama}{MAMA}{Multi-Anode Microchannel Array}
\newacronym{ATLAST}{ATLAST}{Advanced Technology Large Aperture Space Telescope}
\newacronym{PICTURE}{PICTURE}{Planet Imaging Concept Testbed Using a Rocket Experiment}
\newacronym{LITES}{LITES}{Limb-imaging Ionospheric and Thermospheric
Extreme-ultraviolet Spectrograph}
\newacronym{LBT}{LBT}{Large Binocular Telescope}
\newacronym{LBTI}{LBTI}{Large Binocular Telescope Interferometer}
\newacronym{KIN}{KIN}{Keck Interferometer Nuller}
\newacronym{SHARPI}{SHARPI}{Solar High-Angular Resolution Photometric Imager}
\newacronym{IRAS}{IRAS}{Infrared Astronomical Satellite}
\newacronym{HARPS}{HARPS}{High Accuracy Radial velocity Planetary}
\newacronym{hstSTIS}{STIS}{Space Telescope Imaging Spectrograph}
\newacronym{spitzerIRAC}{IRAC}{Infrared Array Camera}
\newacronym{spitzerMIPS}{MIPS}{Multiband Imaging Photometer for Spitzer}
\newacronym{spitzerIRS}{IRS}{Infrared Spectrograph}
\newacronym{CHARA}{CHARA}{Center for High Angular Resolution Astronomy}
\newacronym{wfirst-afta}{WFIRST-AFTA}{Wide-Field InfrarRed Survey
Telescope-Astrophysics Focused Telescope Assets}
\newacronym{GPI}{GPI}{Gemini Planet Imager}
\newacronym{WFIRST}{Roman}{Nancy Grace Roman Space Telescope}
\newacronym{HabEx}{HabEx}{Habitable Exoplanet Observatory Mission Concept}
\newacronym{LUVOIR}{LUVOIR}{Large UV/Optical/Infrared Surveyor}
\newacronym{FGS}{FGS}{Fine Guidance Sensor}
\newacronym{STIS}{STIS}{Space Telescope Imaging Spectrograph}
\newacronym{MGHPCC}{MGHPCC}{Massachusetts Green High Performance
Computing Center}
\newacronym{WISE}{WISE}{Wide-field Infrared Survey Explorer}
\newacronym{ALMA}{ALMA}{Atacama Large Millimeter Array}
\newacronym{GRAIL}{GRAIL}{Gravity Recovery and Interior Laboratory}
\newacronym{jwstNIRCam}{NIRCam}{near-\gls{IR}-camera}
\newacronym{jwstMIRI}{MIRI}{Mid-Infrared Instrument}

\newacronym{AURIC}{AURIC}{The Atmospheric Ultraviolet Radiance Integrated Code} 
\newacronym{FFT}{FFT}{Fast Fourier Transform  }
\newacronym{MODTRAN}{MODTRAN   }{ MODerate resolution atmospheric TRANsmission }
\newacronym{idl}{IDL}{$\textit {Interactive Data Language}$}
\newacronym[sort=NED,description=NASA/IPAC Extragalactic Database]{ned}{NED}{\gls{nasa}/\gls{ipac} Extragalactic Database}
\newacronym{iraf}{IRAF}{Image Reduction and Analysis Facility}
\newacronym{wcs}{WCS}{World Coordinate System}
\newacronym{pegase}{PEGASE}{$\textit{Projet d'Etude des GAlaxies par Synthese Evolutive}$}
\newacronym{dirty}{DIRTY}{$\textit{DustI Radiative Transfer, Yeah!}$}
\newacronym{CUDA}{CUDA}{Compute Unified Device Architecture}
\newacronym{KLIP}{KLIP}{Karhunen-Lo`eve Image Processing}

\newacronym{MSIS}{MSIS}{Mass Spectrometer Incoherent Scatter Radar}
\newacronym{nmf2}{$N_m$}{F2-Region Peak density}
\newacronym{hmf2}{$h_m$}{F2-Region Peak height}
\newacronym{H}{$H$}{F2-Region Scale Height}

\newacronym{isr}{ISR}{Incoherent Scatter Radar}
\newacronym[description=TLA Within Another Acronym]{twaa}{TWAA}{\gls{tla} Within Another Acronym}
\newacronym[plural=SNe, firstplural=Supernovae (SNe)]{sn}{SN}{Supernova}
\newacronym{EUV}{EUV}{Extreme-Ultraviolet }
\newacronym{EUVS}{EUVS}{\gls{EUV} Spectrograph}
\newacronym{F2}{F2}{Ionospheric Chapman F Layer }
\newacronym{F10.7}{F10.7}{ 10.7 cm radio flux [10$^{-22}$ W m$^{-2}$ Hz$^{-1}$]  }
\newacronym{FUV}{FUV}{far-ultraviolet }
\newacronym{IR}{IR}{infrared}
\newacronym{MUV}{MUV}{mid-ultraviolet }
\newacronym{NUV}{NUV}{near-ultraviolet }
\newacronym{O$^+$}{O$^+$}{Singly Ionized Oxygen  Atom }
\newacronym{OI}{OI}{Neutral Atomic Oxygen Spectroscopic State }
\newacronym{OII}{OII}{Singly Ionized Atomic Oxygen Spectroscopic State }
\newacronym{PSF}{PSF}{point spread function}
\newacronym{$R_E$}{$R_E$}{Earth radii [$\approx$ 6400 km]  }
\newacronym{RV}{RV}{radial velocity}
\newacronym{UV}{UV}{ultraviolet }
\newacronym{WFE}{WFE}{wavefront error}
\newacronym{sed}{SED}{spectral energy distribution}
\newacronym{nir}{NIR}{near-infrared}
\newacronym{mir}{MIR}{mid-infrared}
\newacronym{ir}{IR}{infrared}
\newacronym{uv}{UV}{ultraviolet}
\newacronym[plural=PAHs, firstplural=Polycyclic Aromatic Hydrocarbons (PAHs)]{pah}{PAH}{Polycyclic Aromatic Hydrocarbon}
\newacronym{obsid}{OBSID}{Observation Identification}
\newacronym{SZA}{SZA}{Solar Zenith Angle}
\newacronym{TLE}{TLE}{Two Line Element set}
\newacronym{DOF}{DOF}{degrees-of-freedom}
\newacronym{PZT}{PZT}{lead zirconate titanate}
\newacronym{ADCS}{ADCS}{attitude determination and control system}
\newacronym{COTS}{COTS}{commercial off-the-shelf}
\newacronym{CDH}{C$\&$DH}{command and data handling}
\newacronym{EPS}{EPS}{electrical power system}

\newacronym{PCA}{PCA}{principal component analysis}
\newacronym{fwhm}{FWHM}{full-width-half maximum}
\newacronym{RMS}{RMS}{root mean squared}
\newacronym{RMSE}{RMSE}{root mean squared error}
\newacronym{MCMC}{MCMC}{Marcov chain Monte Carlo}
\newacronym{DIT}{DIT}{discrete inverse theory}
\newacronym{SNR}{SNR}{signal-to-noise ratio}
\newacronym{PSD}{PSD}{power spectral density}
\newacronym{NMF}{NMF}{non-negative matrix factorization}

\newacronym{OS}{OS}{observing scenario}
\keywords{Debris disks --- High contrast imaging --- Polarization observations --- Coronagraphs --- Exozodis}

\section{Introduction} \label{sec:intro}
Despite recent advances in high-contrast imaging, circumstellar debris disks around main sequence stars are still poorly understood. Debris disks are composed of planetesimals, predominantly of dust with a small percentage of gas, resulting from successfully formed planetary systems. The analysis and study of debris disks provide valuable insights into the planet formation process and the structure of a planetary system \citep{backman2004debris,hughes2018debris}. 
Our solar system can be broken into the inner hot and warm zodiacal dust, the cool asteroid belt, and the Kuiper belt that form the debris disk of our solar system \citep{wyatt2016insights, ACLR2020}.
\par The boundaries of these populations are sculpted by the gravitation influence of solar system planets, which suggests an indirect technique for detecting planets via gap clearing \citep{stark2008detectability,kennedy2015warm}, a distinct process from planet-driven gas shocks which open gaps in protoplanetary disks \citep{bae2017formation}. 
The size of observed gaps in debris disks will depend heavily on dust composition since, for a fixed planet mass, the size of the gap depends on the transport rates, which is a function of the ratio of radiation pressure to gravitational attraction, commonly known as $\beta$. 
Accurately calculating $\beta$ requires detailed knowledge of the dust grain properties: size, shape, composition, and porosity. {In addition to the gaps, planetary companions are known to induce other features in the debris disks, such as warps, clumps, spirals, and brightness asymmetry \citep{wyatt2008evolution}.}
Resolved, multi-wavelength observations of debris disks reveal general complementary information about the composition and morphology: optical and near-infrared (NIR) observations reveal scattered light by sub-micron and micron-sized dust grains, while observations in infrared (IR) and radio show thermal emission by sub-mm dust grains.  
\par
Although debris disks have been observed {around a hundred stars} in the last two decades, constraining disk properties has not been straightforward. Dust grains' properties contribute in a complicated and often degenerate way to radiative transfer processes, making it quite challenging to disentangle and constrain them individually {using radiative transfer modeling} \citep{krivov2010debris}. Scattered light from debris disks is expected to be linearly polarized due to asymmetries in their structure or scattering/absorption by the dust grains, where the polarization fraction ($p=\sqrt{Q^2+U^2}/I$, where $I$, $Q$, and $U$ are the Stokes parameters, \citep{stokes1852xxx}) as a function of scattering angle {(the angle between the incident wave and the direction of the scattered wave)} is also sensitive to {specific dust grain properties such as composition, size, and distribution}. Thus, when combined, total and polarized {intensity} measurements help constrain the geometrical and scattering properties of the debris disk more than with either one alone \citep[e.g.][]{2020parriaga}. Additionally, polarimetry can improve sensitivity to polarized sources relative to unpolarized star-light, improving the effective contrast ratio \citep{2015Perringpi}.
\par
Polarimetric observations of debris disks have been carried out using the {Advanced Camera for Surveys coronagraph (ACS) in Hubble Space Telescope (HST) and current ground-based high-contrast imaging polarimeters} in optical and NIR wavelength regions \citep{graham2007signature, maness2009hubble, engler2017hip, milli2017hr4796a, Esposito2020, chen2020multiband, hom2020first, crotts2021deep, hull2022polarization}. Using the Gemini Planet Imager (GPI) \citep{macintosh2014first} at the Gemini-South Telescope, \citet{esposito2020debris} conducted a four-year survey of 104 stars and obtained polarization observations of 35 debris disks at NIR wavelengths. In addition to this, the {Spectro Polarimetric High-Contrast Exoplanet REsearch (SPHERE)/Zurich Imaging Polarimeter (ZIMPOL) \citep{beuzit2006sphere,schmid_sphere/zimpol_2018}, the SPHERE/The infrared dual-band imager and spectrograph (IRDIS)\citep{vigan2010photometric}, and the Nasmyth Adaptive Optics System - Near Infrared Imager and Spectrograph (NaCo) at the VLT \citep{witzel2011instrumental}, and the Subaru Coronagraphic Extreme Adaptive Optics (SCExAO) \citep{martinache2016subaru}/High-Contrast Coronographic Imager for Adaptive Optics (HiCIAO) \citep{hodapp2008hiciao} at the Subaru telescope have also been used to image debris disks in polarization at NIR and optical wavelengths to constrain disk properties \citep{milli2017hr4796a,engler2017hip,asensio2016polarimetry}}. For example, polarimetric observations of HR 4796A using Gemini/GPI and VLT/SPHERE/ZIMPOL have helped to place some constraints on its dust grain composition \citep{2020parriaga, 2019milliopt}. {Similarly, using the total and polarized intensity observations through HST/ACS, \cite{graham2007signature} identified distinctions between two approaches of modeling scattering phase functions, Henyey-Greenstein and Mie theory, to model the properties of the debris disk AU Mic. Although polarized observations have proven extremely useful in deriving the disk properties, it is quite challenging to simultaneously obtain the polarized and (unbiased) total intensity observations \citep{esposito2018direct}.}
\par
The upcoming Nancy Grace Roman Space Telescope Coronagraph Instrument \citep{poberezhskiy2022roman, poberezhskiy2021roman,kasdin2020nancy} will facilitate polarimetric observations of debris disks (sensitive to star-planet flux ratios of $\lesssim10^{-8}$) around nearby stars in addition to the high-contrast and high-resolution imaging of exoplanets. The polarimetric module consists of two Wollaston prisms, each producing two orthogonally polarized images ($I_0$, $I_{90}$ and  $I_{45}$, $I_{135}$) that are separated by 7.5''. Polarimetric imaging is available for the narrow field with the Hybrid Lyot Coronagraph (HLC) at 575 nm and the wide field with the Shared Pupil Coronagraph (SPC) at 825 nm. The accuracy requirement in a linear polarization fraction (LPF) measurement per spatial resolution element (2$\times$2 pixels in HLC and 3$\times$3 pixels in SPC) is $<$ 3\%. However, Monte Carlo simulations using uncertainty in calibration, flat fielding, and photometric noise on standards estimate an RMS error in LPF measurement per resolution element to be 1.66\% \citep{mennesson2021roman,zellem2022nancy}.
\par Polarization observations through the Roman Space Telescope will play a vital role in providing constraints on some of the disks already observed by GPI and SPHERE with its complementary observations in optical wavelength regimes in addition to resolving the fainter dust rings much closer (around 1 AU) to nearby stars. One of the potential problems in polarimetry is the errors due to instrumental polarization and crosstalk arising due to the telescope and instrumental optics, which has been very well observed in GPI \citep{Millar-Blanchaer2016SPIE,millar2022polarization}, SPHERE/IRDIS \citep{deBoer2020,vanHolstein2020}, and SPHERE/ZIMPOL \citep{schmid2018sphere}. {To best plan the calibration strategies for Roman-CGI polarimetry observations, it is crucial to characterize the effect of instrumental polarization and crosstalk. Additionally, accurate end-to-end disk polarization observation simulations enable optimized observation planning through Roman-CGI.}
\par
In this work, we describe a process for simulating the polarization observations of debris disks through the Roman Coronagraph Instrument and {demonstrate the instrument's potential to measure linear polarization fraction greater or equal to 0.3 with an uncertainty of less than 0.03. To demonstrate our approach, we generate simulated polarimetric observations of debris disk models around a nearby (early) Sun-like star Epsilon-Eridani ($\epsilon$ Eridani) and HR 4796A, an A0V star harboring an extensively studied debris disk\footnote{We intend to model a disk with similar geometry as HR 4796A, as the scattering properties of the disk considered here are different from what has actually been observed.}. As the main motivation of this paper is to demonstrate the simulations of disks through Roman-CGI, we use disk models that are analogues to the disks around these stars but do not necessarily agree with all the existing multi-wavelength observations.} 
{The paper is organized as follows: \S \ref{sec:model} describes the mathematical model for the observation simulation. The radiative transfer modeling approach of the $\epsilon$ Eridani and HR 4796A disks is described in \S \ref{sec:diskmodel}. Generation of Point Response functions for the HPC and SPC mode is described in \S \ref{sec:Prfs}. \S \ref{sec:emccd} shows the creation of raw EMCCD images. The processing of disks incorporating noise and uncertainty factors from the \gls{OS} simulations is described in \S \ref{sec:os9sim}, and estimations of output polarization fractions are provided in \S \ref{sec:finalpol}. Finally, we provide our discussions and conclusions in \S \ref{sec:conclusion}.}
\\
\section{Modeling Methods}
\label{sec:model}
\begin{figure*}[!ht]
\begin{center}
\includegraphics[width=1\linewidth]{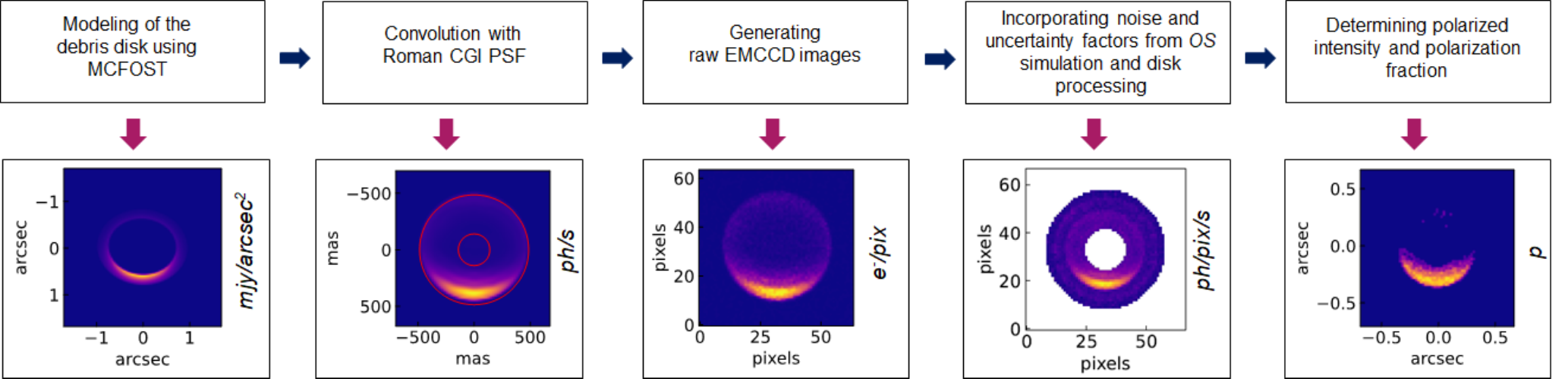}
\caption{The process for simulation of the polarization observations of debris disks through the Roman Coronagraph. The Stokes parameters for the linear polarization are obtained from the radiative transfer modeling of the debris disk using MCFOST. They are then converted into orthogonal polarization components and propagated through the instrument, incorporating various noise sources and instrumental polarization effects. The observable polarization fraction is estimated from the processed disk images.}
\label{mathmodel}
\end{center}
\end{figure*}

To give the reader tangible examples, we will describe the modeling process for two example systems, one so far unresolved in scattered light and one well-known, extensively studied.
Figure \ref{mathmodel} describes the outline of the different steps involved in the simulation of a polarization observation of a debris disk. First, we model the debris disk using the radiative transfer modeling software \href{https://ipag.osug.fr/~pintec/mcfost/docs/html/overview.html}{MCFOST} \citep{Pinte2006, Pinte2009} to obtain the total intensity image and Stokes parameters $Q$ and $U$ images for linear polarization.  {Next, the orthogonal polarization components are convolved with the Roman Coronagraph point response functions (PRFs) obtained using the \href{http://proper-library.sourceforge.net/}{PROPER} \citep{krist2007proper} models run for the HLC and SPC modes. We use \href{http://proper-library.sourceforge.net/}{PROPER}, which combines coronagraphic modes with EMCCD properties directly, instead of higher-level coronagraph models such as FALCO \citep{riggs2018fast} or CGISim. FALCO is mainly used for wavefront sensing and control, and \href{https://sourceforge.net/projects/cgisim/}{CGISim} is not required as we incorporate the EMCCD noise after the convolution.} 
Next, the EMCCD raw images, including the EMCCD gain and noise characteristics, are generated. We then add the speckle and jitter noise using Observing Scenario 9  (``OS9")\footnote{\url{https://roman.ipac.caltech.edu/sims/Coronagraph_public_images.html\#Coronagraph_OS9}} simulations of the Roman Coronagraph for the HLC mode and \gls{OS} 11 (``OS11")\footnote{\url{https://roman.ipac.caltech.edu/sims/Coronagraph_public_images.html\#Coronagraph_OS11_SPC_Modes}} simulations for the SPC mode. These images are processed as conventional CCD images with high read noise (no photon counting) using the ``analog" mode as described in \cite{nemati2020photon}. The final step is the estimation of Stokes parameters, polarization intensity, and total intensity after incorporating the instrumental polarization and polarization crosstalk from the pupil-averaged Mueller matrices\footnote{\url{https://roman.ipac.caltech.edu/docs/Roman-Coronagraph-Optical-Model-Mueller-Matrices-450-to-950nm.pdf}} of the instrument. Each step in the simulation is explained in detail in the following subsections.
\section{Generating disk models using MCFOST}
{To illustrate the different modes of operation of Roman Coronagraph, relevant disk observation of two archetypal debris disks systems: the inner $\epsilon$ Eridani system will be simulated through the \gls{HLC} mode, and the HR 4796A system will be simulated through the wide-field \gls{SPC} mode. Although the inner disk of $\epsilon$ Eridani has never been resolved, we choose to model this system to exemplify the potential of Roman Coronagraph compared to previous coronagraphic instruments.}
\label{sec:diskmodel}
\subsection{$\epsilon$ Eridani}
\label{modeling}
$\epsilon$ Eridani is a star similar to the early Sun, at a distance of 3.2 pc and T*=5100K, M*=0.82\(M_\odot\) and R*= 0.88\(R_\odot\) \citep{di2004vlti, van2007validation, mamajek2008improved}. The outer debris disk around $\epsilon$ Eridani has been resolved at infrared and submillimeter wavelengths \citep{aumann_iras_1985,greaves_dust_1998,macgregor_epsilon_2015,booth_northern_2017}; {the inner disk, however, is currently unresolved, and its structural and grain properties have not yet been constrained \citep{su2017inner,mawet_deep_2018,wolff_hiding_2023}. The debris disk has been typically divided into two components: 1) a warm inner disk, which is sometimes hypothesized as an unresolved excess consisting of two narrow belts (1.5-2 AU and 8-20 AU; \citep{su2017inner}; 3 AU and 20 AU; \citep{backman2008epsilon}) and 2) a resolved cold outer disk (55-80 AU \citep{su2017inner} or 90-110 AU \citep{backman2008epsilon}) imaged with both ALMA \citep{booth2017northern} and other sub-millimeter instruments \citep{backman2008epsilon,greaves2005structure}.}
{We model the inner warm disk with two narrow belts (1.5-2 AU and 8-20 AU) using MCFOST. Dust properties are taken from \cite{su2017inner} and shown in Table \ref{mcfost-para}. We use Mie theory \citep{mie1908beitrage} as the scattering model in MCFOST as it allows for the complete treatment of polarization. As the modeling suggests two separated inner belts, many possible inner structures are possible, which could indicate the presence of companions, and upcoming JWST observations are expected to better constrain the properties of the system; however, the parameters used here provide a physically plausible model to illustrate the sensitivity of the Roman Coronagraph.}
\begin{figure*}[!ht]
\begin{center}
\includegraphics[width=0.9\linewidth]{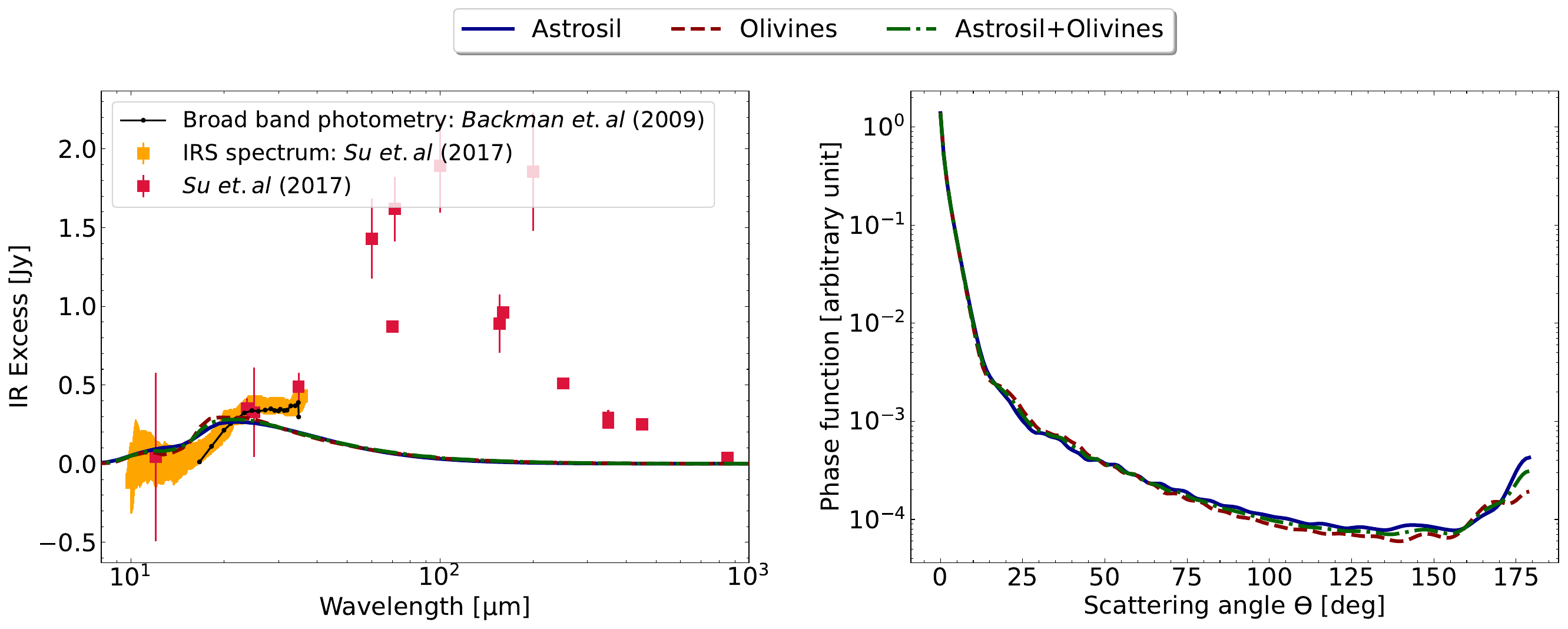}
\caption{Left: IR excess estimated from MCFOST for $\epsilon$ Eridani using astrosilicates (100\%), olivines (100\%), and astrosilicates (50\%)+olivines (50\%) for the inner-most ring and  H20 dominated dirty ice (100\%) for the central ring, compared with observations from \citep{su2017inner} and broadband photometry from \citep{backman2008epsilon}. Right: Corresponding scattering phase function (SPF) at 575 nm (central wavelength of HLC band) obtained from MCFOST for the dust grain compositions. The three dust grain compositions show a reasonable match with the observed IR excess and show similar SPF.}
\label{sed-dust}
\end{center}
\end{figure*}
\par
{The IR excess estimated from the {MCFOST modeled spectral energy distribution (SED)} is compared with the observed Spitzer-IRS spectrum obtained from \cite{su2017inner} and broadband photometry from \cite{backman2008epsilon} as shown in the left panel of Figure \ref{sed-dust}. The three models shown in Figure \ref{sed-dust} use the same parameters from Table \ref{mcfost-para} except for the grain composition for the inner-most ring, either  100\% astrosilicates, 100\% olivine, or astrosilicates (50\%)+ olivine (50\%) to demonstrate that all the three models estimate similar IR excess, SPF. We estimated a higher IR excess than the observed values using 100\% amorphous carbon or 100\% graphite, while 100\% dirty ice shows lower values (not shown here). Among the three dust grain compositions shown in Figure \ref{sed-dust}, we generated our disk model using the parameters in Table \ref{mcfost-para}. The model IR excesses match reasonably well with the observed IR excess at mid-IR wavelengths. As our disk model contains only the two inner rings, we do not attempt to match the SED beyond 25 microns, which would represent the outer ring. This simplification has negligible impact on the scattered visible light inside 1\arcsec~and allows for improved sampling of the circumstellar region of interest.}
The disk in our simulations is modeled with an inclination \textit{i} of 34$^{\circ}$ and position angle (PA) of 266$^{\circ}$ \citep{booth2017northern} for the narrow band filter with a bandpass FWHM of 56.5nm and central wavelength of 575nm. The scattered light and Stokes parameter images are 256 $\times$ 256 pixels in size with a pixel scale of 21.84 mas/pixel.
\begin{table*}[!ht]
\begin{center}
\begin{tabular}{ccc}
\hline
Dust rings                          & \multicolumn{1}{c}{Inner most}  & \multicolumn{1}{c}{Central}  \\ \hline 
Disk extent (AU)               & 1.5-2            & 8-20      \\
Scale Height (AU)             & 0.03 at 1.5 AU              & 0.3 at 8 AU \\
Dust Mass (\(M_\odot\))      & 1.50$\times10^{-12}$       & 1.00$\times10^{-12}$ \\
Minimum grain size-$a_{min}$ ($\mu$m)           & 1  & 1 \\
Maximum grain size-$a_{max}$ ($\mu$m)          & 1000  & 1000 \\
Power-law of grain size distribution        & 3.65   & 3.65    \\
Grain composition     & 100\% astrosilicates       & 100\% "dirty" ice \\ \hline
\end{tabular}
\end{center}
\caption{Parameters used in the MCFOST modelling of $\epsilon$ Eridani from \cite{su2017inner}. As shown in Figure \ref{sed-dust}, olivines (100\%) or astrosilicates(50\%)+olivines (50\%) can also be used for the inner-most ring.}
\label{mcfost-para}
\end{table*}
\par
\begin{figure}[!h]
\centering
\includegraphics[width=0.48\linewidth]{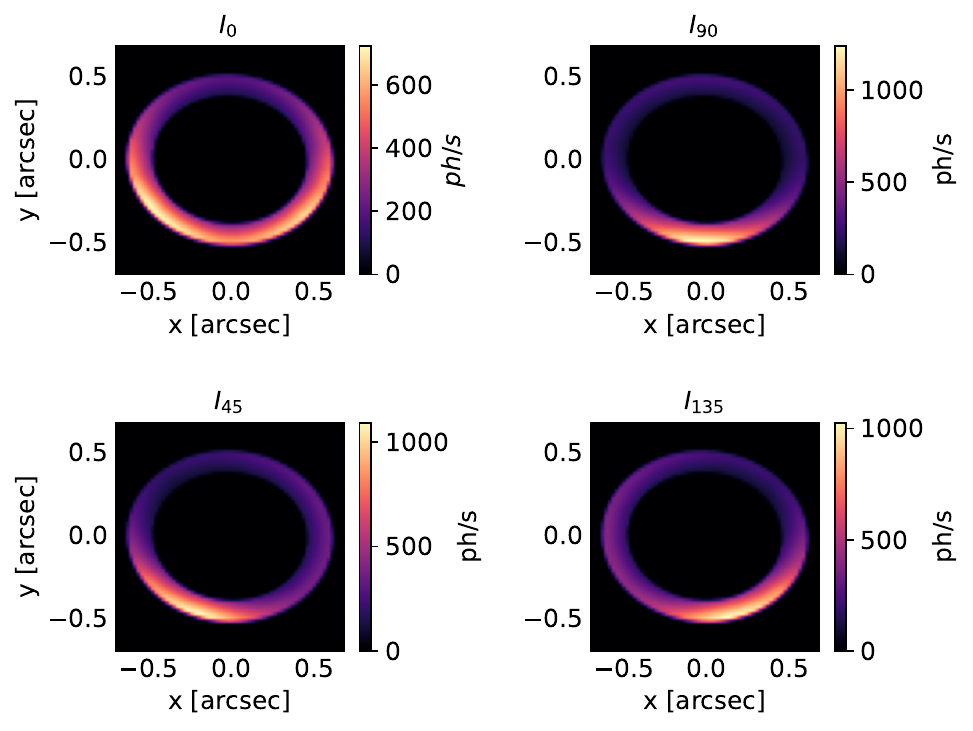}
\includegraphics[width=0.49\linewidth]{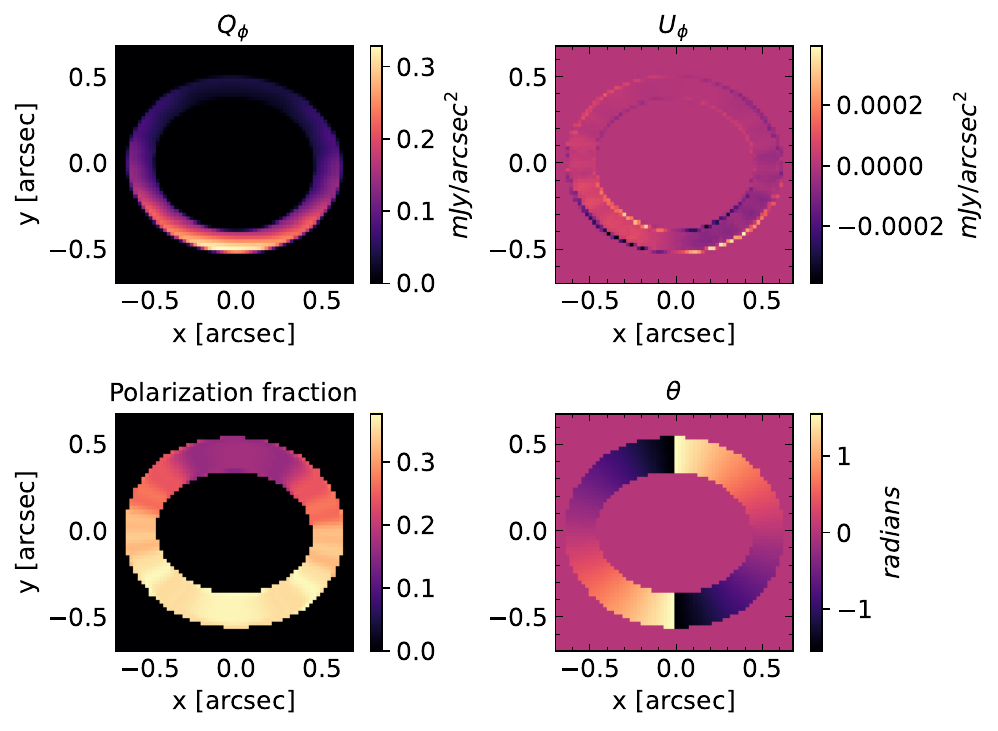}
\caption{Left: The four orthogonal polarization components ($I_0$, $I_{90}$, $I_{45}$, $I_{135}$) in photons/s estimated using the Stokes parameters ($Q$ and $U$) and total intensity obtained from MCFOST for $\epsilon$ Eridani. Right: Input Stokes Azimuthal components $Q_\phi$, $U_\phi$ and corresponding polarization fraction $Q_\phi/I$ and $\theta$ obtained from MCFOST for $\epsilon$ Eridani. A peak polarization fraction of 0.37$\pm$0.01 in one resolution element of 3$\times$3 pixels is estimated in the direction of forward scattering of the disk using the Mie scattering model with 100\% astrosilicates as the dust grain composition.}
\label{pol-input-hlc}
\end{figure}

 The model Stokes parameter images ($I$, $Q$ and $U$, in units of  $W/m^2$) are converted to $Jy$ and further need to be expressed in terms of four orthogonal polarization components for propagation through the instrument. For a perfect instrument (the instrument Mueller matrix is applied later), $I_0$, $I_{90}$, $I_{45}$, $I_{135}$ can be obtained as,
\begin{align}
I_0=\frac{I+Q}{2} \quad
I_{90}=\frac{I-Q}{2} \\
I_{45}=\frac{I+U}{2}  \quad
I_{135}=\frac{I-U}{2}
\end{align}
where $I$ corresponds to the total intensity expressed in $Jy$; these orthogonal polarization components are converted to photons/s using $\zeta$ Pup as the reference star with $V$=2.23 from \gls{OS}9 (``OS9") simulations, estimating 9.3985$\times10^8$ photons/s at the primary mirror of the telescope. The estimated polarization intensities in photons/s at the primary mirror are shown in the left panel of Figure \ref{pol-input-hlc}.
\par
To measure the impact of instrumental noise (see Section \ref{sec:emccd} and \ref{sec:os9sim}), we will compare the final modeled polarization fraction of the disk with the MCFOST-simulated polarization fraction. To estimate the polarization fraction, one can use $p=\sqrt{\frac{Q^2+U^2}{I}}$, but squaring $Q$ and $U$ introduces a systemic bias in low SNR data \citep{schmid2006}. Therefore, we convert the Stokes $Q$ and $U$ images obtained from MCFOST to $Q_\phi$ and $U_\phi$, such that the electric field vector direction (in the polarization) is radially oriented with respect to the central star. Following \cite{schmid2006}, this transformation is given by
\begin{equation}
Q_{\phi}=-Q\cos(2\phi)-U\sin(2\phi)\\  \label{eqn-1}
\end{equation}
\begin{equation}
U_{\phi}=Q\sin(2\phi)-U\cos(2\phi)  \label{eqn-2}
\end{equation}
\begin{equation}
 \phi=\arctan (\frac{x-x^*}{y-y^*})   
\end{equation}
{$x^*$, $y^*$ corresponds to the pixel location of the central star, and $x$, $y$ corresponds to all other pixel locations.} $Q_\phi/I$ gives the polarization fraction as $U_\phi/I$ becomes negligible; the position angle $\theta$ is estimated as $0.5 \arctan(U/Q)$.
 The $Q_\phi$, $U_\phi$, the polarization fraction ($p$) and position angle ($\theta$) are shown in the right panel of Figure \ref{pol-input-hlc}. We estimate the maximum polarized intensity of 0.32 $mJy/arcsec^2$ and the corresponding polarization fraction of 0.37$\pm$0.01 in one resolution element (3$\times$3 pixels) in the direction of forward scattering of the disk using the Mie scattering model. 
 In our $\epsilon$ Eridani MCFOST modeled disks, we scale the polarized intensity and total intensity to a surface brightness of 0.168 $mJy/arcsec^2$ per pixel derived from the expected contrast level of 2 $\times 10^{-8}$ from the non-detection of inner disk of $\epsilon$ Eridani in HST observations from \cite{douglas2024}
\subsection{HR 4796A} \label{sec:SPC}
{HR 4796A is an A0V star with a bright, inclined debris disk well-studied (distance = 71.9$\pm$0.70pc; \citealt{van2007validation,prusti2016gaia}) across the optical and NIR wavelengths in polarization using many instruments, Gemini/GPI, VLT/SPHERE, and VLT/NaCo.} \cite{2009hinkleyspeckle} obtained the first detection of the NIR polarized intensity of the disk at the ansae. {The front and back sides of the disk were later resolved in polarized intensity with GPI \citep{2015Perringpi}. The improved spatial resolution and smaller IWA identified a brightness asymmetry along the front side of the disk.} The data favored an optically thick, geometrically thin model showing a more substantial forward scattering peak at the smallest scattering angles. Using NaCo and SPHERE at the VLT,  \cite{2015millinew,2017millinearir,2019milliopt} detected asymmetry between the northwest and southeast sides of the disk and measured the polarization phase function of the dust in the disk for the first time. These observations were modeled with MCFOST to derive best-fit parameters for {grain properties.} The observed polarization fraction was found to be 0.4$\pm$0.26 (40\%$\pm$ 26\%) at 90\textdegree~scattering angle in the optical band (VBB broadband filter), and the observed averaged polarized phase function was compared with both a Henyey-Greenstein phase function model and a Mie theory model with micron-sized dust grains.
\par
{Furthermore, multi-wavelength polarization NIR observations presented in \cite{2020parriaga} provided consistent polarization fraction measurements to \cite{2009hinkleyspeckle} and  \cite{2015Perringpi} but showed that single grain composition modeling through Mie (spherical grains) and DHS (distributed hollow spheres) \citep{1988MNRAS.234..209J} theories could not reproduce both the observed polarized and total intensity scattering phase functions (SPF) simultaneously. In a different study, \cite{2020chen} used a DHS grain model with a grain composition of silicates (42\%), carbon(17\%), and metallic iron (37\%) to model the VLT/SPHERE H2 total intensity SPF.}
\begin{figure}[!h]
\centering
\includegraphics[width=1\linewidth]{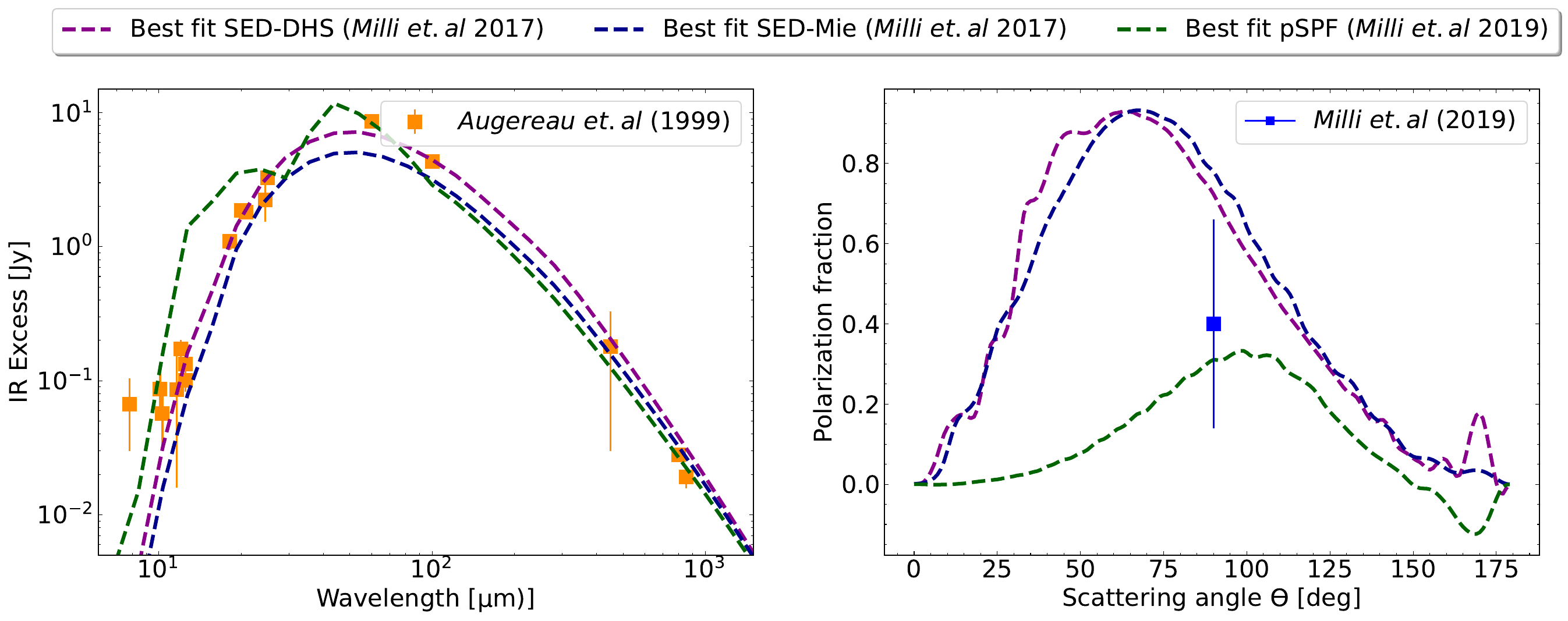}
\caption{Left: IR excesses for HR 4796A generated with MCFOST for the best-fit models from \cite{2017millinearir,2019milliopt}. The observed IR excess is shown from \cite{augereau1999hr}. Right: The polarization fraction at 825 nm (central wavelength of SPC band) obtained from MCFOST for the three different best-fit models from \cite{2017millinearir,2019milliopt}, along with the VLT/ZIMPOL (600-900 nm) measurement of polarization fraction at 90\textdegree scattering angle \citep{2019milliopt}. The models for best-fit SED and best-fit polarized intensity scattering phase function (pSPF) predict different IR excess values and polarization, a tension that Roman polarimetry has the potential to resolve.}
\label{irex-hr4796}
\end{figure}
\par {At present, the geometrical parameters of the disk derived from all the multi-wavelength observations are largely consistent. In contrast, dust grain models poorly match the observed total intensity phase function and the polarization fraction. This mismatch arises from the shape of the grains and the composition, indicating the importance of using more complicated grain models in addition to improved extraction of the total intensity phase function from the observed data \citep{tazaki2019unveiling,arnold2019effect}. Hence, HR 4796A will be one of the crucial targets to be observed through the Roman Coronagraph in polarization and total intensity, which may better inform the existing dust grain models.}
\begin{table*}[!ht]
\begin{center}
\begin{tabular}{cc}
\hline
Parameters & best-fit SED-Mie \\
\hline
Disk extent (AU)               & 60-100                 \\
Scale Height (AU)             & 2               \\
Dust Mass (\(M_\odot\))      & 1.0$\times10^{-6}$        \\
Minimum grain size-$a_{min}$ ($\mu$m)           & 1.77828   \\
Maximum grain size-$a_{max}$ ($\mu$m)          & 10000   \\
Power-law of grain size distribution        & 3.5      \\
Porosity                                    & 0.396 \\
Grain composition     & astrosil (44.15\%)+"dirty" ice (0.66\%)+ carbon (55.18\%)        \\ \hline
\end{tabular}
\end{center}
\caption{Disk properties used in the MCFOST modelling of HR 4796A from \cite{2017millinearir} for best-fit SED using Mie theory.}
\label{mcfost-para-HR4796}
\end{table*}
\par
To simulate the polarization observations of HR 4796A through the SPC mode of the Roman Coronagraph, we use one of the best-fit models as an example disk from \cite{2017millinearir, 2019milliopt} shown in Figure \ref{irex-hr4796}. The IR excesses estimated for three disk models: a) best-fit SED using Mie theory, b) best-fit SED using DHS theory, and c) best-fit polarization fraction profile (pSPF) are compared with the observed values from \cite{augereau1999hr} in the left panel of Figure \ref{irex-hr4796}. The polarization fraction obtained from MCFOST is shown in the right panel of Figure \ref{irex-hr4796} and is compared with the VLT/ZIMPOL measurement at 90\textdegree scattering angle from \citep{2019milliopt}. The best-fit SED models are marginally consistent with the observed polarization fraction.
\par
\begin{figure}[!h]
\centering
\includegraphics[width=0.5\linewidth]{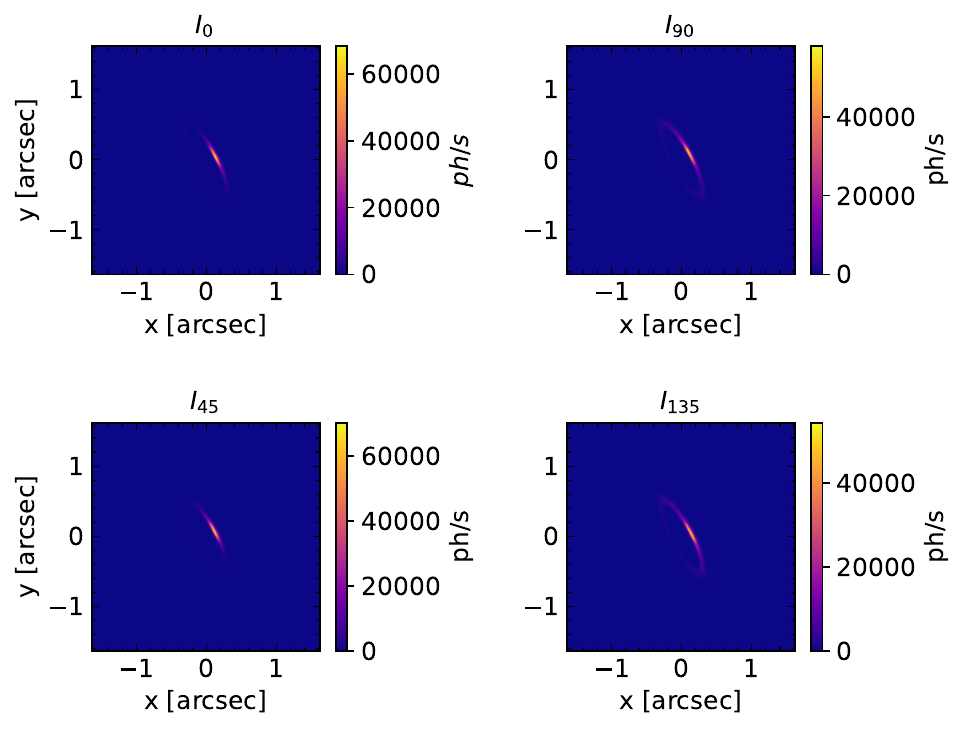}
\includegraphics[width=0.48\linewidth]{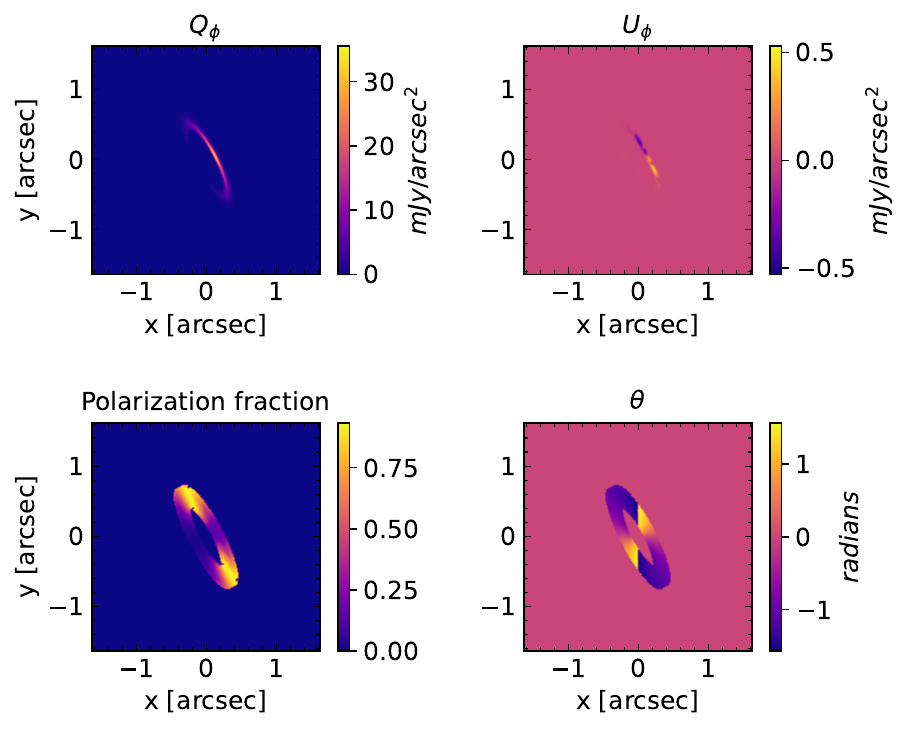}
\caption{Left: The four orthogonal polarization components ($I_0$, $I_{90}$, $I_{45}$, $I_{135}$) in photons/s estimated using the Stokes images ($Q$ and $U$) and total intensity image obtained from MCFOST for HR 4796A. Right: Input Stokes radial components $Q_\phi$, $U_\phi$, corresponding polarization fraction $Q_\phi/I$, and $\theta$ obtained from MCFOST.}
\label{pol-input-spc}
\end{figure}

 We use the disk properties from the best-fit SED model given in Table \ref{mcfost-para-HR4796} for our simulations. The disk is modeled with an \textit{i} of 75.8\textdegree~and PA of 27.7\textdegree~for the broadband filter with a bandpass FWHM of 96.8 nm and a central wavelength of 825.5 nm. The dimensions and pixel scale of the Stokes images are the same as $\epsilon$ Eridani. The Stokes images obtained using MCFOST are converted to orthogonal polarization components in photons/s using $\zeta$ Pup as the PSF reference star (estimating 2.560$\times10^8$ photons/s at the primary mirror of the telescope obtained from \href{https://roman.ipac.caltech.edu/sims/Coronagraph_public_images.html\#Coronagraph_OS11_SPC_Modes}{``OS11''} simulations in the SPC mode of Coronagraph.) The left panel of Figure \ref{pol-input-spc} shows the orthogonal polarization components and the right panel shows $Q_\phi$, $U_\phi$, $p$, and $\theta$ estimated Equation \ref{eqn-1} and \ref{eqn-2}. We estimate a peak polarization fraction of 0.92 $\pm$ 0.01 and a polarized intensity of 35.25 $mJy/arcsec^2$, respectively. 
\par
The orthogonal polarization components of disk models of $\epsilon$ Eridani and HR 4796A shown in the left panel of Figures \ref{pol-input-hlc} and \ref{pol-input-spc}, have to be convolved with the Point Response Function (PRF) of the Roman Coronagraph instrument which is described in the following section.
\section{Generating PRFs for the Roman Coronagraph}
\label{sec:Prfs}
\begin{figure}[!h]
\begin{center}
\fbox{\includegraphics[width=1\linewidth]{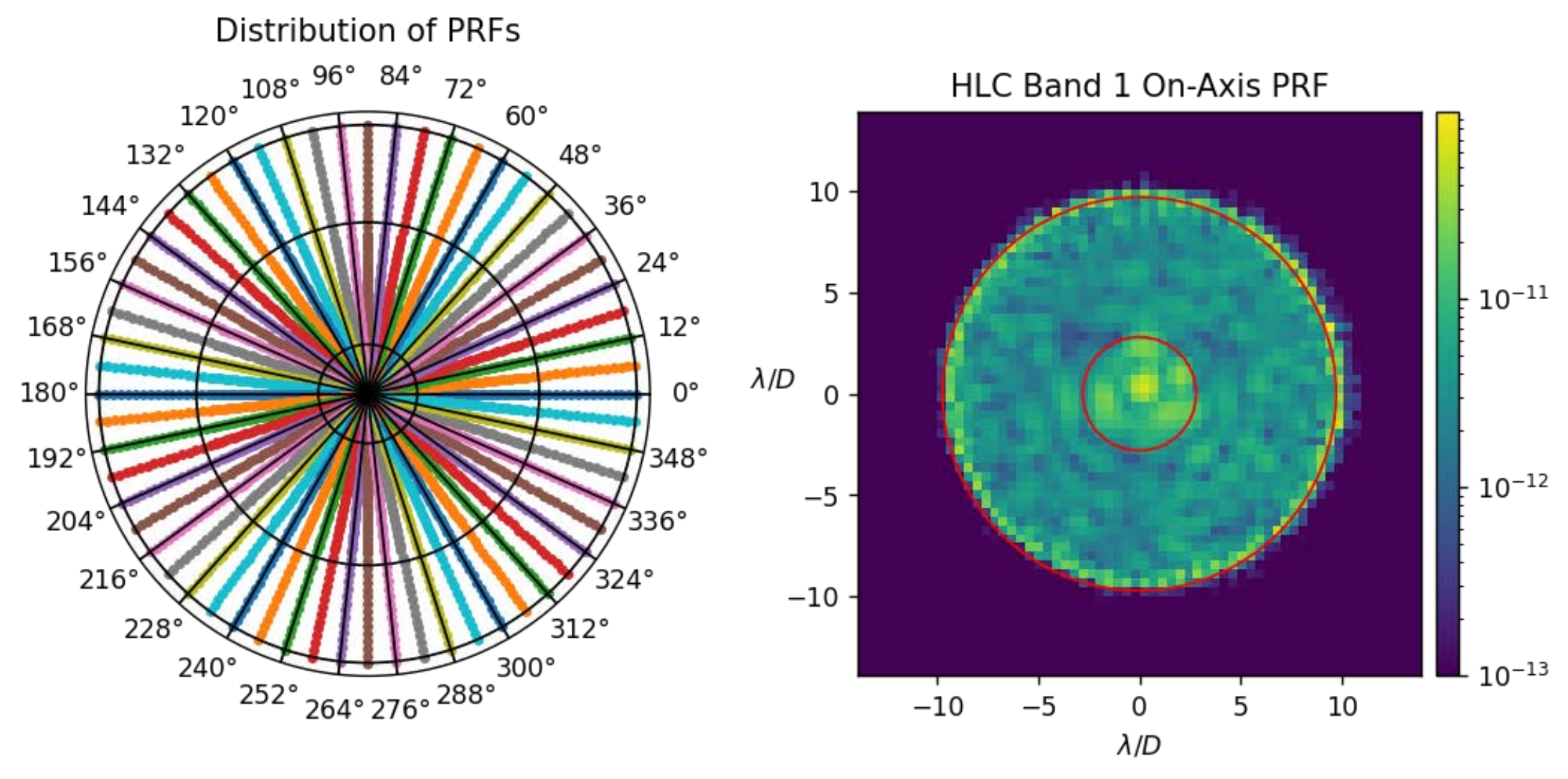}}
\caption{Left: Distribution of PRFs used for the HLC disk simulations along with the on-axis PSF for the HLC mode. The three concentric rings indicate the IWA, OWA, and maximum radial PRF used for each mode. Respectively, these values, in units of $\lambda/D$, are 2.8, 9.7, 15.2 (140.52, 486.83, 762.87 in \textit{mas}). The HLC uses a more dense grid of PRFs to reduce numerical artifacts found in the HLC simulations from the interpolation of the PRFs. {Right: On-axis PRF corresponding to stellar leakage. The wavefronts of each PRF are normalized at the Roman entrance pupil such that the total sum is 1.}}
\label{prfs}
\end{center}
\end{figure}

For each coronagraph mode, a dataset of Point Response Functions (PRFs) is generated using the end-to-end CGI propagation models in \href{https://sourceforge.net/projects/cgisim/files/}{roman-phasec-proper (v1.2.5)} {utilizing \href{http://proper-library.sourceforge.net/}{PROPER} as the back-end propagator.} Note that the term Point Response Function is used instead of Point Spread Function as PSF will often imply a linear and shift-invariant instrument response. Due to the influence of apodizers and the focal plane mask, we do not assume a shift-invariant response, so a standard convolution with a PSF cannot be used to generate simulations of disks at the detector. Instead, the dataset of PRFs for a particular mode is interpolated to the array of pixel coordinates of the disk model. This method reduces the image simulation step of a particular disk to a matrix-vector multiplication, as explained in \cite{milani2020fastersims}. Crucially, by utilizing a large set of PRFs that sample the FOV in the radial and angular coordinates, as shown in Figure \ref{prfs}, this method captures the field dependence of the coronagraph PRF within the image simulation. Additional PRFs extending beyond the OWA capture the scattering contributions from sources extending beyond the nominal FOV.
\par {To form the PRF matrix,} given a wavelength and source offset, a wavefront is propagated through the optical train to simulate a single monochromatic image. Each HLC PRF is an incoherent sum of seven wavelengths within the band 1 filter centered at 575nm. The SPC PRFs each use five wavelengths within the band 4 filters centered at 825nm. The polarization aberration setting is set to the mean of all polarization states (polaxis=10 within the \href{http://proper-library.sourceforge.net/}{PROPER} models) as we incorporate the polarization effects from the Mueller matrix of the Roman Coronagraph in the final step of simulations. The wavefront of each PRF is normalized to have a total amplitude of 1 at the entrance pupil of the Roman aperture. 
\begin{figure}[!ht]
\begin{center}
\fbox{\includegraphics[width=1\linewidth]{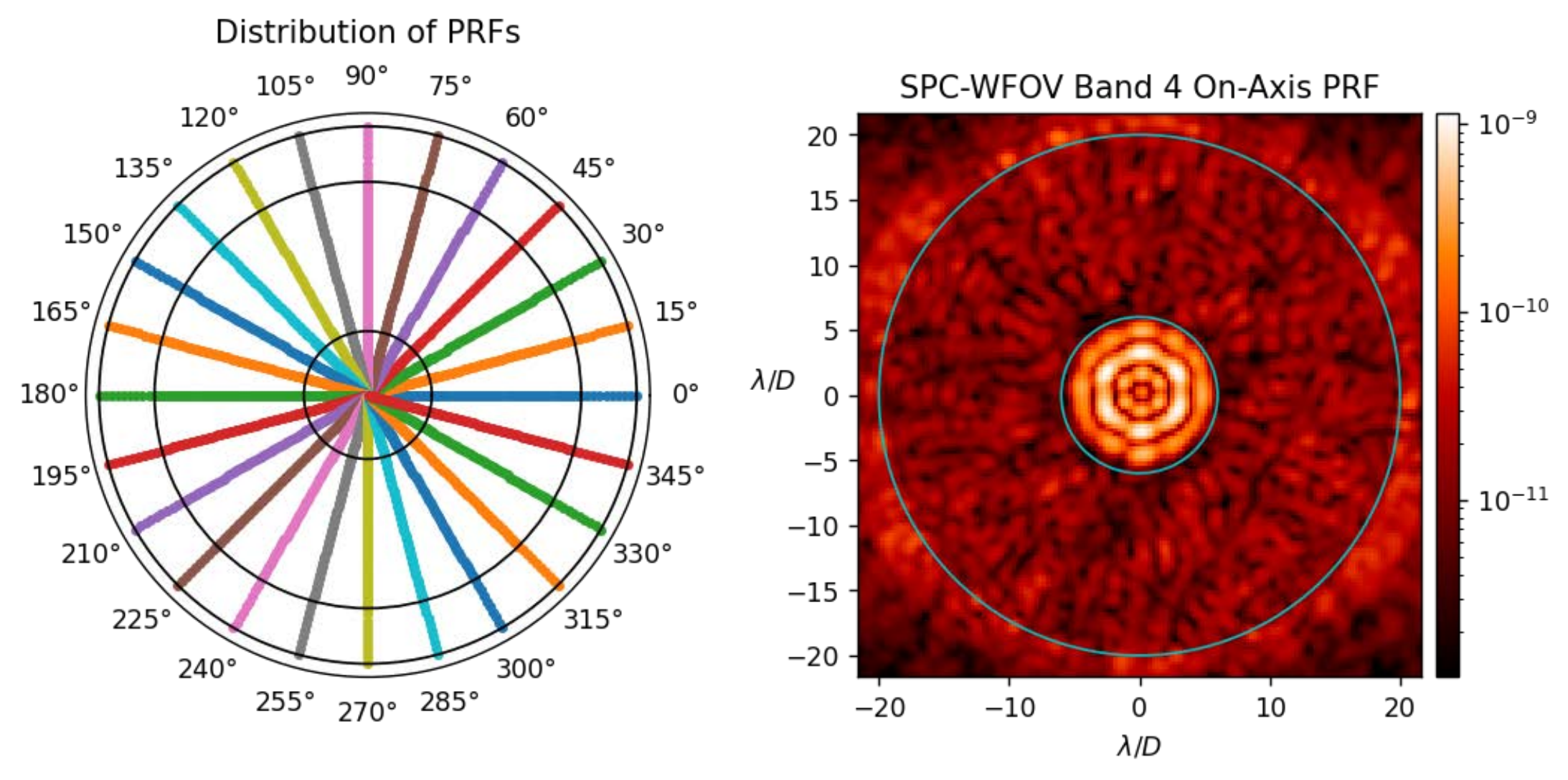}}
\caption{Left: Distribution of PRFs used for the SPC-WFOV disk simulations along with the on-axis PRF. The three concentric rings indicate the IWA, OWA, and maximum radial PRF. These values, in units of $\lambda/D$, are 6, 20, 25.2 (432.06, 1440.21, 1814.66 in \textit{mas}), respectively. {Right: On-axis PRF corresponding to stellar leakage. Once again, the wavefronts of each PRF are normalized at the Roman entrance pupil such that the total sum is 1.}}
\label{spc-prfs}
\end{center}
\end{figure}
\par
\begin{figure*}[!h]
\begin{center}
\includegraphics[width=0.49\linewidth]{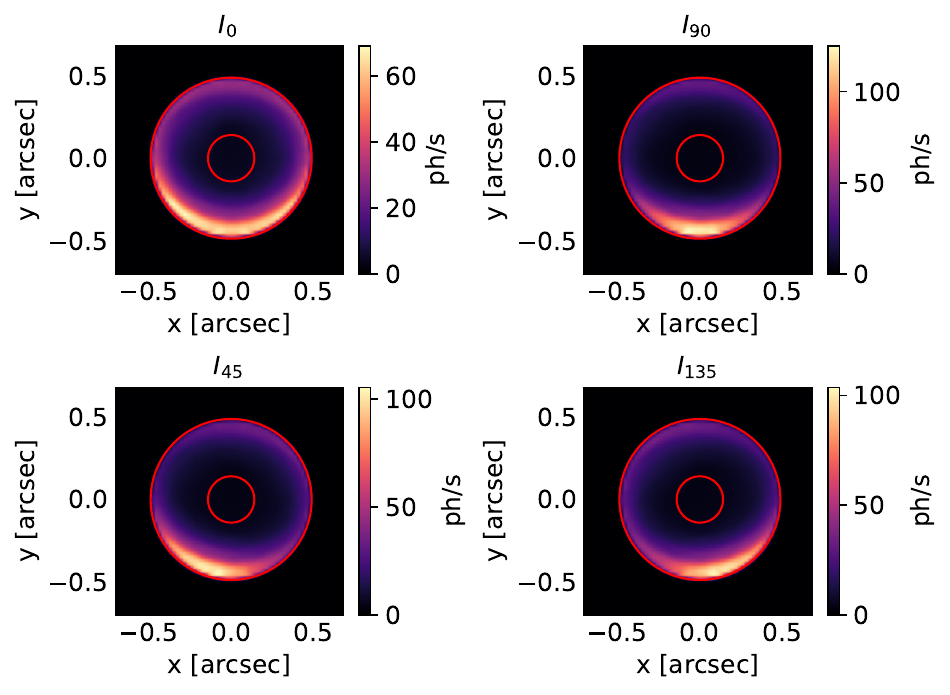}
\includegraphics[width=0.48\linewidth]{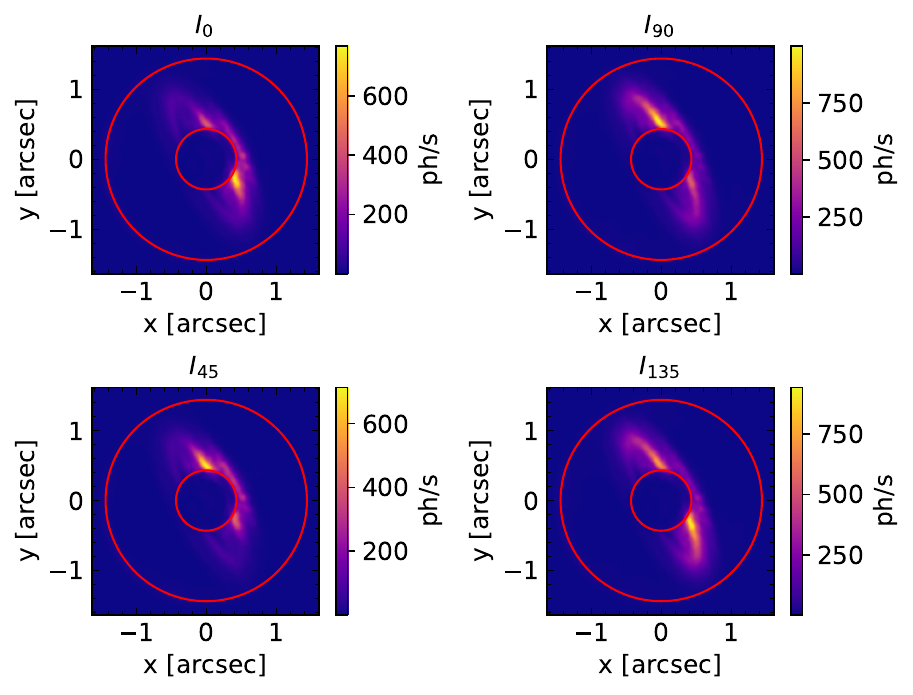}
\caption{Left: The convolved orthogonal polarization components in $ph/s$ through the HLC mode for $\epsilon$ Eridani. The IWA (140.52 \textit{mas}) and OWA (486.83 \textit{mas}) are marked using concentric red circles. Right: The convolved orthogonal polarization components for HR 4796A through the SPC mode. The IWA (432.06 \textit{mas}) and OWA (1440.21 \textit{mas}) are marked using concentric red circles}
\label{polintsim}
\end{center}
\end{figure*}

 The convolved disk images of the orthogonal polarization components $I_0$, $I_{90}$, $I_{45}$, and $I_{135}$ are shown in Figure \ref{polintsim} for $\epsilon$ Eridani (left panel) and HR 4796A (right panel) with the inner working angles (IWAs) and outer working angles (OWAs) of the coronagraphs overlaid in red. The disk and PSF pixel scales are maintained to be consistent during the convolution.
\section{Generating raw EMCCD images}
\label{sec:emccd}
\begin{figure}[!h]
    \centering
    \includegraphics[width=0.49\linewidth]{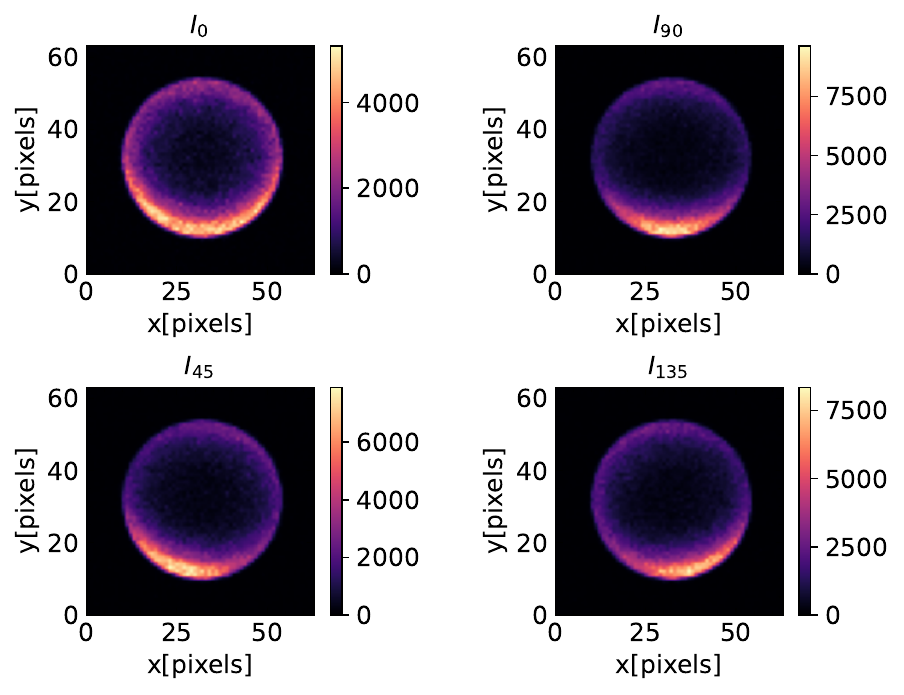}
    \includegraphics[width=0.5\linewidth]{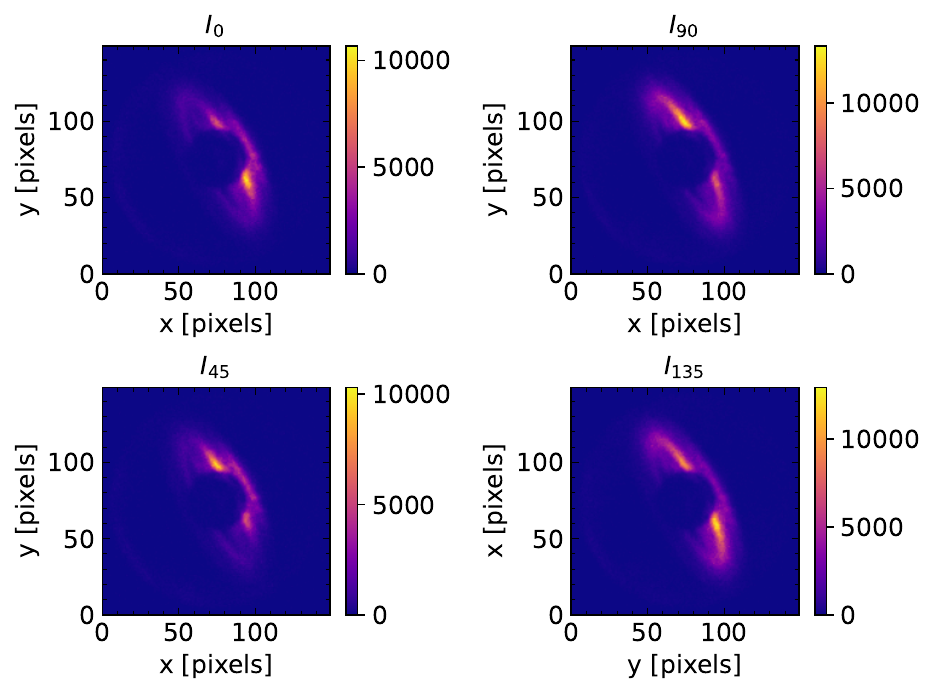}
    \caption{Left: One of the raw EMCCD frames in $electons/pixel$ from the stack of 50 is shown for $\epsilon$ Eridani Right: One of the raw EMCCD frames for HR 4796A.}
    \label{emccd-frames}
\end{figure}

The Roman Coronagraph will use a back-illuminated electron-multiplying CCD sensor (\href{https://www.teledyneimaging.com/en/aerospace-and-defense/products/sensors-overview/ccd/ccd201-20/}{e2v CCD201-20}) consisting of 1024$\times$ 1024 pixels of 13 $\mu$m in size. It can be operated in low gain ($<$1000) and high gain ($>$1000) modes. We use \href{https://github.com/wfirst-cgi/emccd_detect}{\texttt{emccd\_detect}} \citep{nemati2020photon} to simulate the raw EMCCD images from the convolved disk images. 
{A stack of 50 EMCCD frames is simulated for each orthogonal polarization component with an exposure time of 5s/frame for $\epsilon$ Eridani and 1s/frame for HR 4796A, a gain of $\le$ 200 incorporating bias of 700$e^-$, dark current of 0.0028 $e^-/pix/s$, and read noise of 100 $e^-$ and also incorporate photon noise.} Figure \ref{emccd-frames} shows one of the frames at the EMCCD for all four orthogonal polarization components for $\epsilon$ Eridani and HR 4796A, respectively.  
\section{Incorporating noise and uncertainty factors from \gls{OS} simulations and disk processing}
\label{sec:os9sim}
\gls{OS} simulations are the simulated science images created by the integrated modeling team at NASA-JPL using the most recent version of the observation strategy. They include end-to-end Structural Thermal Optical Performance (STOP) models of the Roman observatory, coronagraph masks, diffraction, wavefront control, detector noise, and jitter. 
The observing sequence starts with a slew from a Wide Field Instrument (WFI) target to the CGI reference star ($\zeta$ Pup), followed by 30 hours of observatory settling time before the coronagraphic observations as described in \cite{ygouf2021roman}. {In the most recent \gls{OS}11 simulations, there are four coronagraphic observation cycles with time between the 2nd and 3rd cycle for the dark hole maintenance. Assuming the dark hole was previously dug and required only minor modifications, each observation cycle begins with observing the reference star ($\zeta$ Pup) for 45 minutes, followed by 100 minutes of target star (47 Uma) observations at each of 4 rolls, alternating between ‐13 \textdegree~and +13\textdegree~ twice for a total of 400 minutes on target per cycle. The reference star is imaged for 45 minutes again at the end of the cycle. During the four cycles, the reference star is observed six times for a total of 4.5 hours and the target star for 26.67 hours.}
\par The Roman observatory's STOP model is run for a specified timestep to simulate the aberrations and pupil shifts during each observation cycle. Next, the Jitter model produces the RMS jitter over a specified period. Then, the Low order Wavefront Sensing (LOWFS) model is run to generate the Deformable Mirror (DM) correction patterns and is fed to the \href{https://sourceforge.net/projects/cgisim/files/}{roman-phasec-proper} diffraction model of the observatory. For each timestep, \href{https://sourceforge.net/projects/cgisim/files/}{roman-phasec-proper} produces complex values of speckle electric fields for four different polarizations (which can be either used individually or two orthogonal polarizations are added for the case of unpolarized source). Finally, these speckle images are propagated through the EMCCD model to incorporate detector noise and uncertainties. 
Thus, these simulations incorporate all optical aberrations, pointing jitter, DM thermal drifts, polarization aberrations, and EMCCD noise characteristics. These simulations produce data sets with and without noise and also with and without optical model uncertainty factors (MUFs). 
\par
\begin{figure*}[!ht]
\begin{center}
\includegraphics[width=0.485\linewidth]{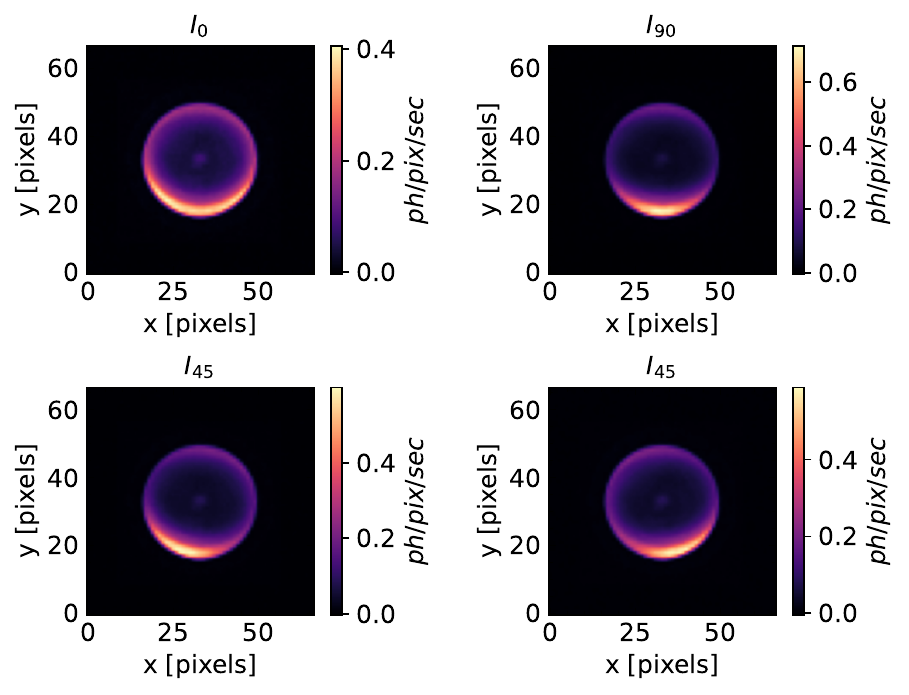}
\includegraphics[width=0.48\linewidth]{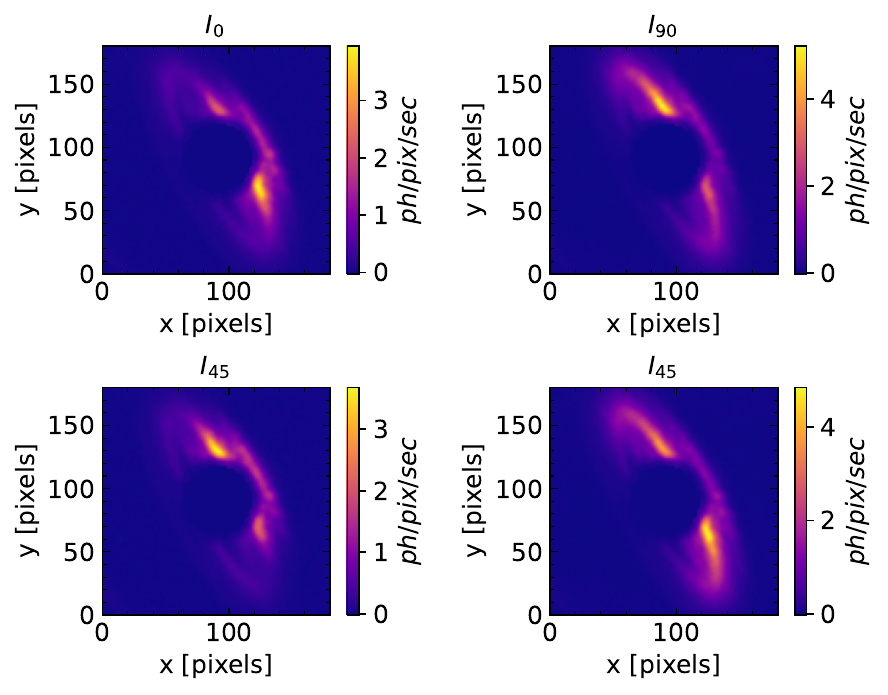}
\caption{Orthogonal polarization components obtained using the ``analog" mode \citep{nemati2020photon} incorporating all the noise components and optical model uncertainty factors (MUFs) from  \gls{OS} simulations. Left: $\epsilon$ Eridani, Right: HR 4796A.}
\label{photon-count}
\end{center}
\end{figure*}

{In our simulations for the HLC mode, the raw EMCCD images of the disks are rotated to the corresponding roll angles following the steps in the observing sequence, and all the noise components and optical model uncertainty factors (MUFs) from the \href{https://roman.ipac.caltech.edu/sims/Coronagraph_public_images.html\#Coronagraph_OS9}{``OS9''} are added. The speckle field images we use in our simulations are [14375, 67, 67] in dimensions, where each speckle field image is obtained for an exposure time of 5s following the \gls{OS} time sequence. Similarly, for the SPC mode \href{https://roman.ipac.caltech.edu/sims/Coronagraph_public_images.html\#Coronagraph_OS11_SPC_Modes}{``OS11''} time series speckle field images are added to the raw EMCCD images to incorporate the speckle noise where the speckle field image has dimensions of [1830, 181, 181] generated for an exposure time of 1s. There are two modes of disk processing to generate the final CCD image as described in the \gls{OS} simulations: ``photon counting" mode with low read noise for gain$>$1000 and ``analog" mode with high read noise for gain$<$1000. We use the ``analog" mode (corresponding to conventional CCD image processing as our disk targets have high SNR) as we use the gain$<$1000 to simulate all the orthogonal polarization components as shown in Figure \ref{photon-count}. In the \gls{OS} simulations, the target star used is 47 Uma with $V=5.4$, and in our simulations, our disk target host stars $\epsilon$ Eridani has $V=3.73$, and HR 4796A has $V=5.744$. We are currently not scaling the speckle fields according to the brightness of our host stars and use the speckle fields generated by the \gls{OS} simulations as our disks are brighter than the speckle noise. We scaled the $I_0$ of  $\epsilon$ Eridani disk a hundred times fainter and processed with the \gls{OS} simulations to understand the speckle noise level. Figure \ref{fainterdisk} in Appendix \ref{faint-disk} shows $\epsilon$ Eridani processed in the analog mode and the corresponding speckle field noise images.}
\section{Estimating polarized intensity and polarization fraction}
The final step in the simulation process is to estimate the Stokes parameters and polarization fraction using the processed disk images of the four orthogonal polarization components. The ($Q_{out}$, $U_{out}$) and total intensity ($I_{out}$) are calculated using
\begin{eqnarray}
I_{out}=I_0+I_{90}; \quad 
Q_{out}=I_0-I_{90}  \\
U_{out}=I_{45}-I_{135}
\label{stokes-eqn}
\end{eqnarray}
The instrumental polarization effects, or shifts in Stokes parameters, introduced due to the telescope and instrument optics are represented as a Mueller matrix \citep{keller2002instrumentation}. The field-independent Mueller matrix for the Roman Coronagraph is provided by the modeling team
\footnote{\url{https://roman.ipac.caltech.edu/sims/Coronagraph_inst_param_data_more.html}} for wavelengths from 450nm to 950 nm. 
\label{sec:finalpol}
\subsection{$\epsilon$ Eridani}
We estimated the averaged Mueller matrices for the HLC band and obtained corrected output Stokes parameters, $Q_{cor}$ and $U_{cor}$ as
\begin{eqnarray}
Q_{cor}&=&-0.0092-0.99Q_{out}+0.99\times10^{-6}U_{out}  \\
U_{cor}&=&0.99U_{out}     
\end{eqnarray}
The instrumental polarization, 0.92\%, and polarization rotation $\rm 0.99\times 10^{-6}$ obtained from the instrument Mueller matrix are within the expected measurement error and can be easily calibrated. The Lu-Chipman decomposition of the Mueller matrices in \cite{doelman2023falco} shows that the negative $Q_{out}$ is obtained due to a mirror and a retarder in the optical path. The $Q_{cor}$ and $U_{cor}$ are converted to $Q_\phi$ and $U_\phi$ using Equations \ref{eqn-1} and \ref{eqn-2}. The \href{https://roman.ipac.caltech.edu/sims/Coronagraph_public_images.html\#Coronagraph_OS9}{``OS9''} repository consists of normalized off-axis PRFs for the Roman Coronagraph, which includes losses from the masks but not from reflections, filters, and Quantum Efficiency (QE). We convolved the flux of the reference star $\zeta$ Pup with the off-axis PSFs to determine the Zero Point (ZP) magnitude (16.11) to correct for the instrument throughput. 
The polarized intensity $Q_{\phi}$ and total intensity images for $\epsilon$ Eridani are shown in the left panel of Figure \ref{output-pol} along with $p$ and $\theta$. We estimate the peak value of $Q_\phi$ as 0.26 $\rm mJy/arcsec^2$, $I_{out}$ as 0.78 $\rm mJy/arcsec^2$, and a peak polarization fraction of 0.33$\pm$0.01 in one resolution element. The peak value of the input polarization fraction shown in the right panel of Figure \ref{pol-input-hlc} is 0.37$\pm$0.01. Thus, we have successfully recovered the input polarization fraction within the measurement error $<$ 3\% after incorporating all the noise sources for a more realistic $\epsilon$ Eridani inner disk.
\subsection{HR 4796A}
The averaged Mueller matrices for the SPC band are used to obtain the corrected output Stokes parameters, $Q_{cor}$, and $U_{cor}$ as,
\begin{eqnarray}
Q_{cor}&=&-0.005-0.99Q_{out}+3.07\times10^{-5}U_{out}\\ U_{cor}&=&-4.307\times 10^{-5}Q_{out}+0.9047U_{out}. 
\end{eqnarray} \\
$Q_{cor}$, and $U_{cor}$ are then converted into $Q_\phi$ and $U_\phi$ using Equation \ref{eqn-1}. The off-axis PRFs from the \gls{OS}11 repositories are used to estimate the Zero-Point magnitude (18.46), using the reference star $\zeta$ Pup for converting ph/pix/sec to their corresponding fluxes.
\begin{figure*}[!ht]
\begin{center}
\includegraphics[width=0.485\linewidth]{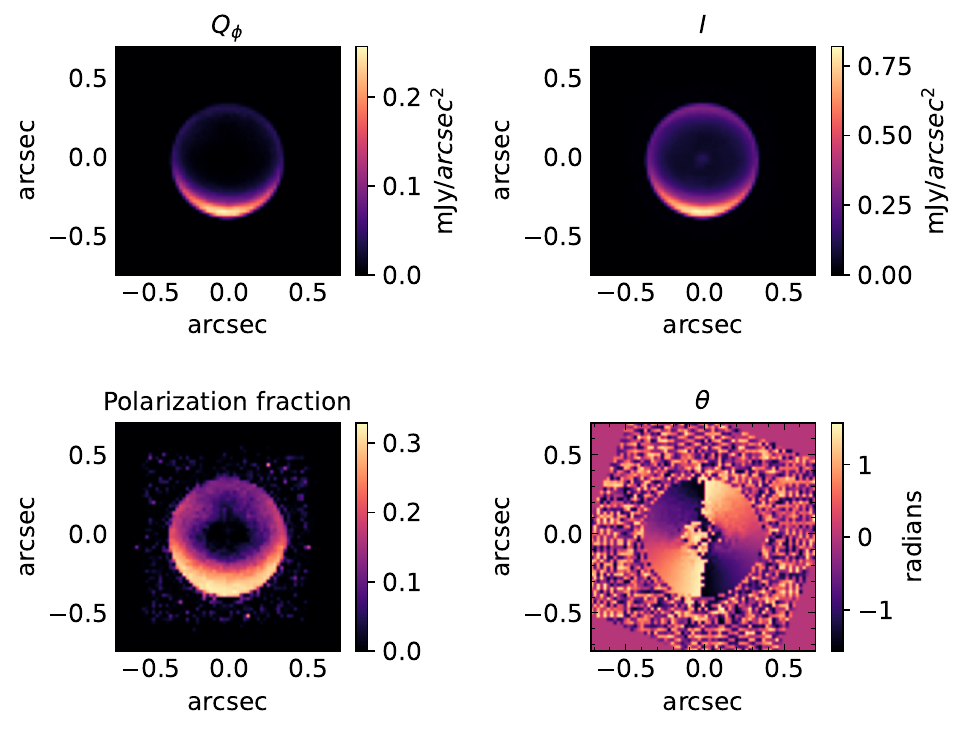}
\includegraphics[width=0.48\linewidth]{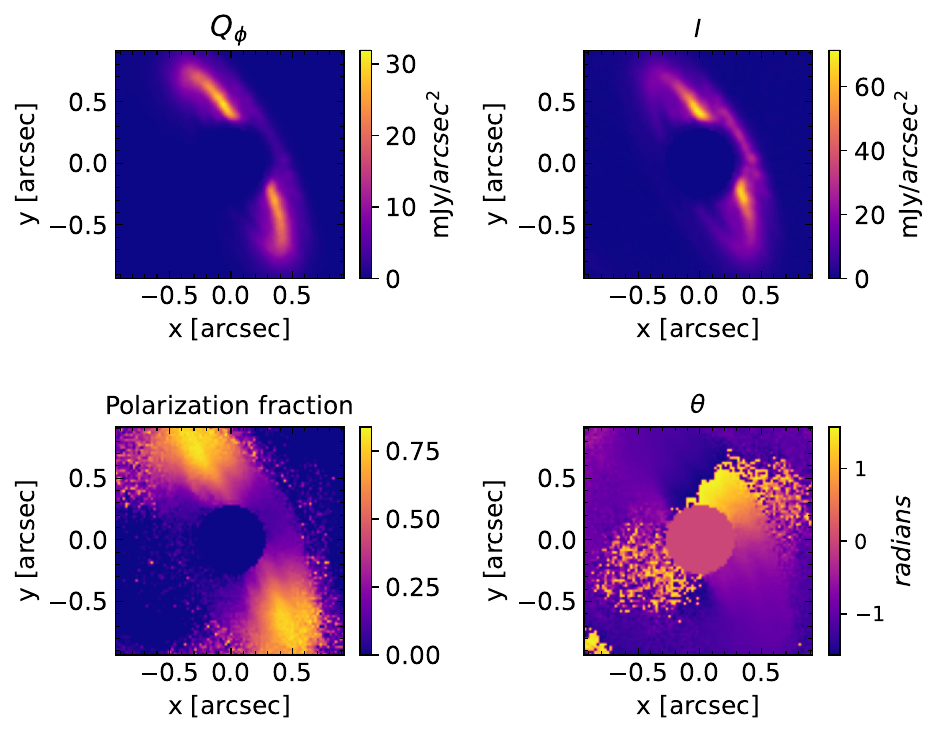}
\caption{The output polarized intensity ($Q_\phi$), total intensity, polarization fraction, and $\theta$ for $\epsilon$ Eridani (left-panel) in HLC mode and HR 4796A (right-panel) in SPC mode.}
\label{output-pol}
\end{center}
\end{figure*}
The output polarized intensity and polarization fraction are shown in the right panel of Figure \ref{output-pol}. We estimate the peak value of $Q_\phi$ as 32.29 $\rm mJy/arcsec^2$, total intensity as 79.03 $\rm mJy/arcsec^2$, and a peak polarization fraction of 0.80$\pm$0.03 in one resolution element. The retrieved polarization fraction for HR 4796A varies on the order of 0.08-0.10 (8-10)\% compared to the input polarization fraction. To investigate this discrepancy, We estimated the polarization fraction after each simulation step. The polarization fraction estimated after the convolution showed a difference of ~0.08-0.10 (8-10)\% with the input polarization fraction (shown in Appendix \ref{pol-diff}), which may be a systematic bias causing reduction of the peak brightness of the disk and can be addressed with accurate forward modeling. Thus, it should be an important part of developing the Roman CGI polarization calibration pipeline. However, we have demonstrated that the polarization observations of HR 4796A through Roman-CGI will help accurately measure the polarization fraction and, hence, may help place better constraints on the dust properties.
\section{Summary and Discussion} 
\label{sec:conclusion}
This work quantifies the expectation that high-contrast polarimetric observations through the Roman Coronagraph can potentially measure the polarization fraction $>$ 0.3 with an uncertainty of 0.03.
\begin{enumerate}
 \item We have developed and presented a pipeline to simulate the polarization observations through the HLC and SPC mode of the Roman Coronagraph instrument. The simulations incorporate detector noise, speckle noise, optical model uncertainty factors (MUFs), and instrumental polarization effects. 
\item {We used MCFOST to model two debris disks, $\epsilon$ Eridani and HR 4796A, and propagated the orthogonal polarization components through instrument simulation tools. We retrieved the peak polarization intensity and the peak polarization fraction from these simulations.}
\item {For simulating $\epsilon$ Eridani through the HLC mode, using astrosilicates as the dust composition, we recovered the input polarization fraction of 0.33$\pm$0.01 at the forward scattering peak after incorporating instrumental polarization and crosstalk.}  
\item Through the SPC mode, we simulated polarization observations of HR 4796A using the best-fit SED parameters derived from ground-based observations in the optical and NIR. We recovered a peak polarization fraction of 0.80$\pm$0.01 after incorporating polarization effects from the Roman Coronagraph.
\item We find a difference of $\sim$0.03-0.10 (3-10)\% in the output polarization fraction with the input for both of the disks processed using HLC and SPC mode after performing the convolution. This indicates a systematic reduction of the peak brightness of the disk, which must be addressed with accurate forward modeling. The difference between the input and the output polarization fraction is due to the strong PSF smearing effect, which is higher for HR4796A as it's a narrower and sharper ring compared to $\epsilon$ Eridani considered here.
\item For the two disks used in our simulations, we obtained sufficiently high SNR with an exposure time of $\sim$ 250s (5s$\times$50 frames) and hence may not require the target acquisition time of $\sim$ 26 hours used in the Observing Scenario simulations. Future modeling and simulation efforts are required to derive the optimal exposure times for Roman disk targets. 
\end{enumerate}
As a technology demonstration \citep{kasdin2020nancy}, the coronagraph is no longer bound by scientific requirements. This work, however, validates that the Roman Coronagraph design meets the science requirement developed early in the design process to: ``map the linear polarization of a circumstellar debris disk that has a polarization fraction greater or equal to 0.3 with an uncertainty of less than 0.03" \citep{douglas_wfirst_2018}. 

This study focused on developing and validating a simulation pipeline for Roman Coronagraph polarimetric observations and demonstration of recovering the polarization fraction from processed disks without considering polarization aberrations \citep{millar2022polarization} and performing ``photon counting". The pipeline for the simulated polarization observations of $\epsilon$ Eridani and HR 4796A is publicly available \citep{Anche_polsim_2023}.
Future work will incorporate the effects of polarization aberrations from the telescope and the coronagraph, perform photon counting, and compare different post-processing methods (e.g., \gls{KLIP}\citep{soummer2012detection} and \gls{NMF}\citep{ren2018non}) for disk extraction, and ultimately assess the retrieval of disk geometric and grain properties. 
\section{Acknowledgments}.
Portions of this work were supported by the WFIRST Science Investigation team prime award \#NNG16PJ24 and the Arizona Board of Regents Technology Research Initiative Fund (TRIF). JA is supported by a NASA Space Technology Graduate Research Opportunity. The authors would like to thank Dr. Bruce Macintosh, Dr. John Krist, and Dr Kate L Su for their support and useful discussions.
\software
{\href{https://www.astropy.org/}{astropy}, 
\href{https://github.com/wfirst-Coronagraph/emccd_detect}{EMCCD detect},
\href{https://ipag.osug.fr/~pintec/mcfost/docs/html/overview.html}{MCFOST},
\href{https://github.com/seawander/nmf_imaging}{$nmf\_imaging $}},
\href{https://github.com/cpinte/pymcfost}{pymcfost}, 
\href{https://pysynphot.readthedocs.io/en/latest/}{pysynphot},
\href{https://scipy.org/}{scipy},
\href{https://pandas.pydata.org/}{pandas}
\href{https://proper-library.sourceforge.net/}{PROPER}
\appendix
\section{Discrepancy in the polarization fraction after convolution}
\label{pol-diff}
\begin{figure*}[!ht]
\begin{center}
\includegraphics[width=0.9\linewidth]{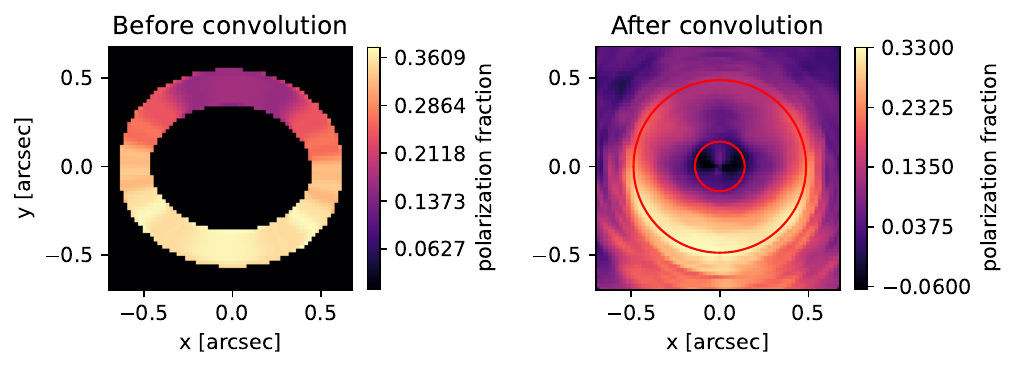}
\includegraphics[width=0.9\linewidth]{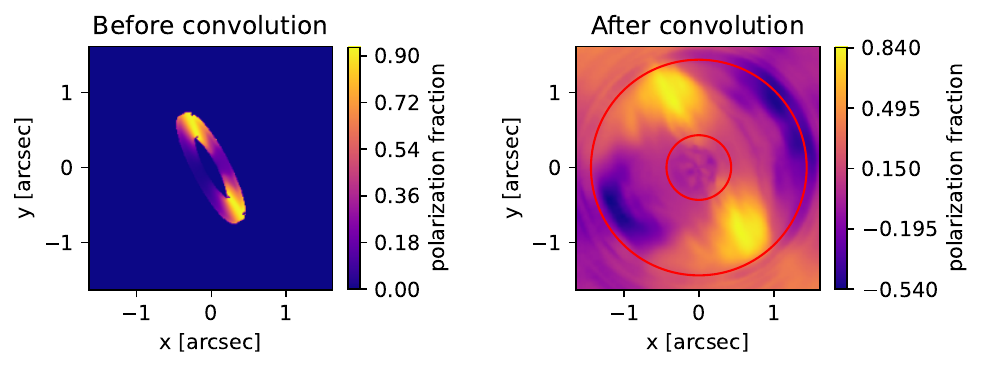}
\caption{The difference in the polarization fraction seen before and after the convolution is shown for $\epsilon$ Eridani (top panel) and HR 4796A (bottom panel).}
\label{poldiff}
\end{center}
\end{figure*}

The polarization fractions estimated for $\epsilon$ Eridani and HR 4796A before and after convolving with the Roman PRFs are shown in Figure \ref{poldiff}. The difference of $\sim$ 3\% and $\sim$ 10\% is observed in the case of $\epsilon$ Eridani and HR 4796A, respectively.
\section{Fainter disk injection}
\begin{figure*}[!h]
\begin{center}
\framebox{\includegraphics[width=1\linewidth]{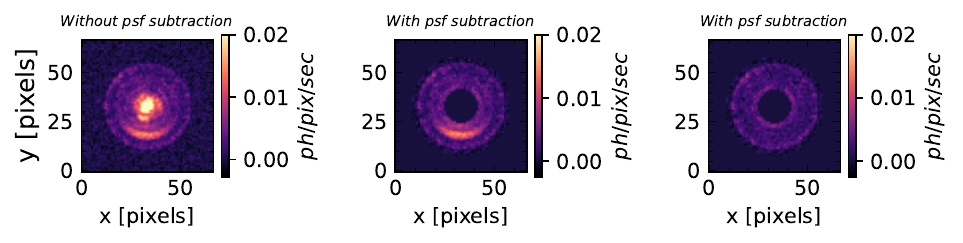}}
\caption{Left: The $I_0$ component of $\epsilon$ Eridani is scaled a hundred times fainter and added to the speckle field noise. Center: The $I_0$ component is processed with the speckle field images.  Right: Speckle field image without the $I_0$ injected.}
\label{fainterdisk}
\end{center}
\end{figure*}

 In our simulations, we are currently not scaling the speckle fields according to the brightness of the host stars $\epsilon$ Eridani and HR 4796A and use the speckle fields generated by the \gls{OS} simulations as our disks are brighter than the speckle noise. We scaled the orthogonal polarization component $I_0$ by 100 times fainter and processed with the \gls{OS}9 simulations to understand the level of speckle-noise fields as shown in Figure \ref{fainterdisk}. The left panel shows the speckle field noise from the \gls{OS}9 simulations added to the $I_0$ component. The mid-panel shows the $I_0$ after the PSF subtraction and the last panel shows the PSF subtracted speckle field image without the $I_0$ component injected.  
\label{faint-disk}
\bibliography{ref,esd_refs}
\bibliographystyle{aasjournal}



\end{document}